\documentclass[12pt]{article}
\usepackage{graphics,cite,amssymb,epsfig,float}
\usepackage[usenames,dvips]{color}
\usepackage{citesort}

\makeatletter

\@addtoreset{equation}{Section}
\makeatother

\def\lsim{\;\raise0.3ex\hbox{$<$\kern-0.75em\raise-1.1ex\hbox{$\sim$}}\;}
\def\gsim{\;\raise0.3ex\hbox{$>$\kern-0.75em\raise-1.1ex\hbox{$\sim$}}\;}

\def\ben{\begin{enumerate}}  \def\een{\end{enumerate}}
\def\bit{\begin{itemize}}    \def\eit{\end{itemize}}
\def\beq{\begin{equation}}   \def\eeq{\end{equation}}
\def\ba{\begin{array}}       \def\ea{\end{array}}
\def\bea{\begin{eqnarray}}   \def\eea{\end{eqnarray}}
\def\nn{\nonumber}

\def\nl{\newline}

\begin{document}

\begin{titlepage}
\renewcommand{\thefootnote}{\fnsymbol{footnote}}
\setcounter{footnote}{0}

\begin{flushright}
LPT Orsay 08-34 \\
\end{flushright}

\begin{center}
\vspace{3cm}
{\Large\bf Phenomenology of the General NMSSM with \\ 
Gauge Mediated Supersymmetry Breaking} \\
\vspace{2cm}
{\bf U. Ellwanger, C.-C. Jean-Louis, A.M. Teixeira} \\
Laboratoire de Physique Th\'eorique\footnote{Unit\'e mixte de Recherche
-- CNRS -- UMR 8627} \\
Universit\'e de Paris XI, F-91405 Orsay Cedex, France
\vspace{2cm}
\end{center}

\begin{abstract}
We investigate various classes of Gauge Mediated Supersymmetry Breaking
models and show that the Next-to-Minimal Supersymmetric Standard Model
can solve the $\mu$-problem in a phenomenologically acceptable way.
These models include scenarios with singlet tadpole terms, which are
phenomenologically viable, e.g., in the presence of a small Yukawa
coupling $\lsim 10^{-5}$. Scenarios with suppressed trilinear $A$-terms
at the messenger scale lead naturally to light CP-odd scalars, which
play the r\^ole of pseudo $R$-axions. A wide range of parameters of
such models satisfies LEP constraints, with CP-even Higgs scalars below
114 GeV decaying dominantly into a pair of CP-odd scalars.
\end{abstract}

\end{titlepage}

\renewcommand{\thefootnote}{\arabic{footnote}}
\setcounter{footnote}{0}

\section{Introduction} 

The mediation of supersymmetry breaking to the observable sector via
supersymmetric gauge interactions (GMSB) has already been proposed
during the very early days of super\-symmetric model building
\cite{early1,earlyoraif}. The essential ingredients of this class of
models are a sequestered sector containing a spurion or a
dynamical superfield $\widehat{X}$, whose $F$-component $F_X$ does not
vanish (there could exist several such fields). In addition, a
messenger sector $\widehat{\varphi}_i$ exists, whose fields have a
super\-symmetric mass $M$, but a mass splitting between its
scalar/pseudoscalar components due to its coupling to $F_X$. They carry
Standard Model gauge quantum numbers such that the messengers couple to
the Standard Model gauge supermultiplets. Possible origins of
supersymmetry breaking in the form of a nonvanishing $F_X$ component
can be O'Raifeartaigh-type models \cite{earlyoraif}, models based on
no-scale supergravity \cite{des,ue1995} or Dynamical Supersymmetry
Breaking \cite{dsb,dsbnmssm,iss}.

If supersymmetric gauge interactions would be the only interactions
that couple the visible sector with the messenger/sequestered sector,
the phenomenologically required $\mu$ and $B\mu$ terms of the Minimal
Supersymmetric Standard Model (MSSM) would be difficult to generate.
The simplest solution to this problem is the introduction of a gauge
singlet superfield $\widehat{S}$ and a superpotential including the
$\lambda\widehat{S}\widehat{H}_u\widehat{H}_d$ term, which has been
used in early globally \cite{nmssm1} and locally supersymmetric
\cite{nmssm2} models. 

Let us point out a possible connection between gravity mediated
supersymmetry breaking and GMSB-like models \cite{des,ue1995}: standard
gravity mediated supersymmetry breaking within the MSSM requires
Giudice-Masiero-like terms (depending on the Higgs doublets) in the
K\"ahler potential \cite{gm} in order to generate the $\mu$ and $B\mu$
terms (see \cite{hmz} for a possible 5-dimensional origin of such
terms). Given a possible source for such terms, one can replace the
Higgs doublets by the messengers of GMSB models and proceed as in the
usual analysis of gauge mediation. The advantage of such models is that
{\it no other} gravity mediated source of supersymmetry breaking as
scalar or gaugino soft masses is required; such sources of
supersymmetry breaking are frequently absent in higher dimensional
setups. On the other hand, the solution of the standard $\mu$-problem
for the Higgs doublets still requires the introduction of a singlet
$\widehat{S}$. Then one is also led to the scenario considered in this
paper, the Next-to-Minimal Supersymmetric Standard Model (NMSSM) with
gauge mediated supersymmetry breaking.

In order to generate a sufficiently large vacuum expectation value of
the scalar component $S$ of $\widehat{S}$ (and hence a sufficiently
large effective $\mu$ term $\mu_{eff} = \lambda \left<
S\right>$), the singlet superfield $\widehat{S}$ should possess
additional Yukawa interactions with the messenger/sequestered sector.
Then, an effective potential for $S$ with the desired properties can be
radiatively generated.

Note that the so-called singlet tadpole problem \cite{tadpole} is
absent once the original source of supersymmetry breaking is of the
$F$-type \cite{des,neme,ue1995}. On the contrary, singlet tadpole
diagrams can now generate the desired structure of the singlet
effective potential \cite{des,ue1995}, triggering a VEV of $S$. If the
singlet couples at lowest possible loop order to the
messenger/sequestered sector such that tadpole diagrams are allowed, a
mild version of the singlet tadpole problem reappears, since the
coefficients of the corresponding terms linear in $S$ are typically too
large. This milder problem can be solved under the assumption that the
involved Yukawa coupling is sufficiently small -- however, it does not
need to be smaller than the electron Yukawa coupling of the Standard
Model (see below).

In the meantime, quite a large number of models involving GMSB and at
least one gauge singlet, that generates an effective $\mu$ term, have
been studied \cite{dgp,gmnm1,gr,gmnm2,dgs,gkr}. They differ in the
particle content of the messenger/sequestered sector, and include
sometimes more than one gauge singlet superfield.

The purpose of the present paper is the investigation of a large class
of models obtained after integrating out the messenger/sequestered
sector (including possibly heavy singlet fields). It is assumed
that the remaining particle content with masses below the messenger
scale $M$ is the one of the NMSSM.

The couplings and mass terms of the NMSSM are obtained under the
following assumptions:\nl
-- no interactions between the Higgs doublets ${H}_u$, ${H}_d$ and the
messenger/sequestered sector exist apart from supersymmetric gauge
interactions; then no MSSM-like $\mu$ or $B\mu$ terms are generated
after integrating out the messenger/sequestered sector;\nl
-- the gauge singlet superfield $\widehat{S}$ has Yukawa interactions
with the messenger/sequestered sector. As a result, various soft terms
and $\widehat{S}$-dependent terms in the superpotential can be
generated after integrating out the messenger/sequestered sector.

Under the only assumption that the original source of supersymmetry
breaking is $F_X$ and that the messengers have a mass of the order $M \gsim
\sqrt{F_X}$, superspace power counting rules allow to estimate the
maximally possible order of magnitude of the generated masses and
couplings.

In general, these masses and couplings will comprise nearly all
possibilities consistent with gauge invariance (see Section 2), leading
to the general NMSSM. However, many of these mass terms and couplings
can be much smaller, or absent, than indicated by the power counting
rules (but never larger), if the corresponding diagrams involve high
loop orders, small Yukawa couplings, or are forbidden by discrete or
(approximate) continuous symmetries.

In the next Section, we will parametrize the mass terms and couplings
of the general NMSSM, and estimate their (maximally possible)
radiatively generated order of magnitude with the help of superspace
power counting rules. Section 3 is devoted to a phenomenological
analysis of three different scenarios, which are defined by particular
boundary conditions for the NMSSM parameters at the messenger scale, and
Section 4 contains our conclusions.

\section{Results of superspace power counting rules}

The class of models investigated in this paper is defined by a
superpotential
\beq\label{2.1}
W=\lambda\widehat{S}\widehat{H}_u\widehat{H}_d
+\frac{\kappa}{3}\widehat{S}^3 +
\widetilde{W}(\widehat{S},\widehat{X},\widehat{\varphi}_i,\dots) 
+\dots\ , \eeq
where
$\widetilde{W}(\widehat{S},\widehat{X},\widehat{\varphi}_i,\dots)$
denotes the couplings of $\widehat{S}$ to the messenger/sequestered
sector, and we have omitted the standard Yukawa couplings of
$\widehat{H}_u$ and $\widehat{H}_d$. No MSSM-like $\mu$-term is assumed
to be present.
Due to a coupling $\widehat{X}\widehat{\varphi}_i\widehat{\varphi}_i$
in $\widetilde{W}$, a non-vanishing $F_X$-component
\beq\label{2.2}
F_X = m^2
\eeq
induces a mass term
\beq\label{2.3}
\frac{1}{2} m^2 \left(A_{\varphi_i}^2 + A_{\varphi_i}^{*\ 2} \right)
\eeq
which gives opposite contributions to the squared masses of the real
and imaginary components of the scalar components of the messengers
$\widehat{\varphi}_i$. Since we assume no direct couplings of
$\widehat{S}$ to $\widehat{X}$, this constitutes the only original
source of supersymmetry breaking.

After integrating out the messenger/sequestered sector, the remaining
effective action for the light superfields $\widehat\Phi$ (the fields
$\widehat{S},\widehat{H}_u,\widehat{H}_d,\dots$
of the NMSSM) is necessarily of the form
\beq\label{2.4}
\sum_i c_i \int d^4\theta f_i(D_\alpha, \overline{D}_{\dot{\alpha}},
\widehat\Phi, \overline{\widehat\Phi},
\widehat{X}, \overline{\widehat{X}})\ ,
\eeq
where the relevant terms are obtained after the replacement of at least
one superfield $\widehat{X}$ by its $F$-component $F_X$.  The maximally
possible orders of magnitude of the coefficients $c_i$ can be obtained
by dimensional analysis: if a function $f_i$ is of a mass dimension
$[M]^{d_f}$, the corresponding coefficient $c_i$ has a mass dimension
$[M]^{2-d_f}$. As long as $d_f \geq 2$ (which will be the case), $c_i$
will typically depend on the mass of the heaviest particle running in
the loops to the appropriate power, and subsequently we identify this
mass $M$ with a unique messenger scale $M_{mess}$.

We are aware of the fact that models exist where the $c_i$ depend on
several mass scales $M_i$; however, it is always trivial to identify a
mass scale $M$ such that $c_i$ are bounded from above by $M^{2-d_f}$.
Also, in the particular case $d_f=2$, $c_i$ can involve large
logarithms; these depend on whether the VEV $F_X$ is ``hard'' (i.e.
generated at a scale $\Lambda$ much larger than $M$) or ``soft'', i.e.
generated by a potential involving terms of the order of $M$. In the
first case, logarithms of the form $\ln(\Lambda^2/M^2)$ can appear in
$c_i$.

In the present situation (no interactions between the Higgs doublets
${H}_u$, ${H}_d$ and the messenger/sequestered sector) possible
supercovariant derivatives $D_\alpha, \overline{D}_{\dot{\alpha}}$
inside $f_i$ do not lead to terms that would otherwise be absent; for
this reason we will omit them in our analysis. (Here, we will not
discuss the radiatively generated gaugino masses and scalar masses for
the gauge non-singlets, but concentrate on the NMSSM specific effects.)
To lowest loop order we can use the underlying assumption that only the
singlet superfield $\widehat{S}$ has direct couplings to the
messenger/sequestered sector (however, see Fig.~1 below). The first
terms that we will investigate are then of the form
\beq\label{2.5}
\sum_i c_i \int d^4\theta f_i( \widehat{S}, \overline{\widehat{S}},
\widehat{X}, \overline{\widehat{X}})\ .
\eeq

Below we list all relevant terms with this structure.
Given an expression of the form (\ref{2.5}), the generated $S$- and
$F_S$-dependent terms can be obtained by the replacements
\beq\label{2.6}
\widehat{X}=M+\theta^2 m^2, \qquad \widehat{S}=S+\theta^2 F_S\ .
\eeq
Due to the coupling $\widehat{X}\widehat{\varphi}_i
\widehat{\varphi}_i$, the supersymmetry conserving mass $M$ of the
messengers $\widehat{\varphi}_i$ can be identified with the value of
the scalar component of $\widehat{X}$. Loop factors like
$(16\pi^2)^{-1}$ and model dependent Yukawa couplings are not
explicitly given, but we indicate the powers of $m$ (which follow from
the powers of $F_X$) and $M$ (which follow from dimensional analysis).

The possible operators $f_i$ and the corresponding contributions to the
scalar potential are then given by:
\bea\label{2.7}
\widehat{S}\overline{\widehat{X}}+h.c.:& & m^2F_S+h.c.\\\label{2.8}
\widehat{S}\overline{\widehat{X}}\widehat{X} +h.c.:& & \frac{m^4}{M} S
+m^2 F_S +h.c.\\\label{2.9}
\widehat{S}\overline{\widehat{S}}{\widehat{X}}+h.c.:& &
\frac{m^2}{M} (S F_S^*+h.c.) + F_S F_S^*\\\label{2.10}
\widehat{S}\overline{\widehat{S}}{\widehat{X}}\overline{\widehat{X}}:& &
\frac{m^4}{M^2}S S^* + \frac{m^2}{M} (S F_S^* +h.c.) +F_S F_S^*\\
\label{2.11} \widehat{S}\widehat{S}\overline{\widehat{X}}+h.c.:& &
\frac{m^2}{M} (S F_S +h.c.)\\
\label{2.12} \widehat{S}\widehat{S}{\widehat{X}}\overline{\widehat{X}} 
+h.c.:& & \frac{m^4}{M^2} S^2 + \frac{m^2}{M} S F_S +h.c.
\eea
Operators with higher powers of $\widehat{X}$ or 
$\overline{\widehat{X}}$ do not generate new expressions, and operators
with higher powers of $\widehat{S}$ or $\overline{\widehat{S}}$
generate negligible contributions with higher powers of $M$ in the
denominator (recall that we are assuming $M \gsim m$).

The terms $\sim F_S F_S^*$ in (\ref{2.9}) and (\ref{2.10}) only account
for a correction to the wave function normalization of the superfield
$\widehat{S}$, which can be absorbed by a redefinition of
$\widehat{S}$. The remaining terms can be written as an effective
superpotential $\Delta W$ and additional contributions  $\Delta
V_{soft}$ to the soft terms of the general NMSSM. To this end, the
terms $S F_S^* +h.c.$ in (\ref{2.9})  and (\ref{2.10}) have to be
rewritten using the expression derived from the superpotential
(\ref{2.1}):
\beq\label{2.13}
F_S^* = \lambda{H_u}{H_d} +{\kappa}{S}^2 + \dots
\eeq
where the dots stand for terms of higher order in the loop expansion.
We parametrize the effective superpotential $\Delta W$ and the soft
terms $\Delta V_{soft}$ of the general NMSSM in agreement with SLHA2
conventions \cite{slha2}:
\bea\label{2.14}
\Delta W &=& \mu'\widehat{S}^2 + \xi_F \widehat{S}\ ,\\ \label{2.15}
\Delta V_{soft} &=& m_S^2\left|S\right|^2 +(\lambda A_\lambda S H_u H_d
+\frac{1}{3} \kappa A_\kappa S^3 + m_S'^2 S^2 +\xi _S S +h.c.)\ .
\eea
Then the expressions (\ref{2.7}) to (\ref{2.12}) lead to
\bea\label{2.16}
\mu' &\sim& \frac{m^2}{M}\ ,\\ \label{2.17}
\xi_F &\sim& m^2\ ,\\ \label{2.18}
m_S^2 &\sim& \frac{m^4}{M^2}\ ,\\ \label{2.19}
A_\lambda = \frac{1}{3}A_\kappa &\sim& \frac{m^2}{M}\ ,\\ \label{2.20}
m_S'^2 &\sim& \frac{m^4}{M^2}\ ,\\ \label{2.21}
\xi_S &\sim& \frac{m^4}{M}\ .
\eea

Next, within the class of models defined by the superpotential
(\ref{2.1}), there exist the diagrams shown in Fig.1 which generate
terms in $\Delta V_{soft}$ which are not included in the list 
(\ref{2.16}) -- (\ref{2.21}). The corresponding operators and soft
terms (after the replacement of $F_{H_u}$ and $F_{H_d}$ by their tree
level expressions) are given by
\bea\label{2.22}
\widehat{H}_u \overline{\widehat{H}}_u \widehat{X}+h.c. & \to 
\frac{m^2}{M} H_u F^*_{H_u} & \to \Delta A_\lambda = \Delta A_{t}
\sim \frac{m^2}{M}\ ,\\ \label{2.23}
\widehat{H}_u \overline{\widehat{H}}_u \widehat{X}
\overline{\widehat{X}} &\to 
\frac{m^4}{M^2} H_u H^*_u &\to \Delta m_u^2 \sim \frac{m^4}{M^2}\ ,
\eea
together with analogous expressions with $H_u$ replaced by $H_d$ (and
$A_{t}$ by $A_{b}$). 

\begin{figure}[ht]
\vskip 0.5cm
\begin{center}
%\begin{picture}(11,8)
\includegraphics[scale=0.64]{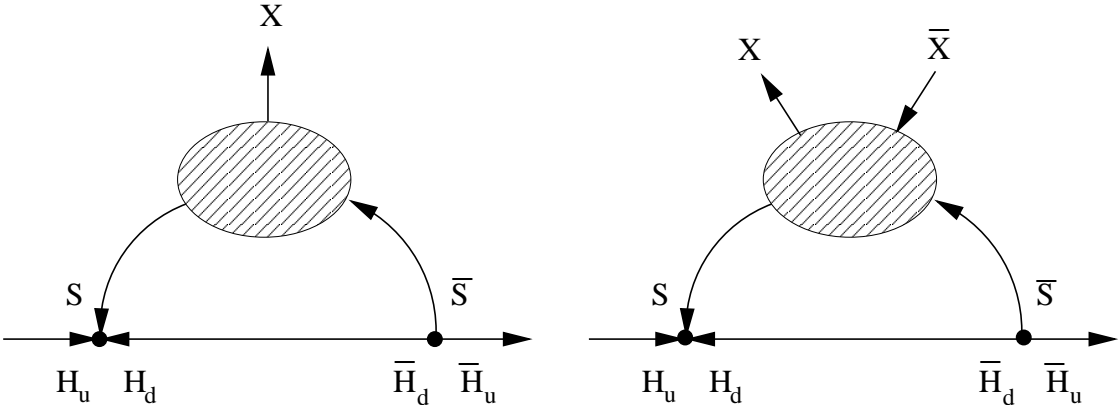}
%\end{picture}
%\vspace*{1cm}
\caption{Superfield diagrams which generate the operators (\ref{2.22})
and (\ref{2.23}) (omitting, for simplicity, the ``hats'' on top of the
letters denoting the superfields.)}
\label{1.1f}
\end{center}
\vspace*{5mm}
\end{figure}

Similar expressions are also generated by i)~the replacement
of the shaded bubbles in Fig.1 by the effective operators (\ref{2.9})
and (\ref{2.10}) (which generate the soft terms (\ref{2.18}) and
(\ref{2.17})), and ii)~the Renormalization Group (RG) evolution of
$A_\lambda$, $A_{t}$, $A_{b}$, $m_u^2$ and $m_d^2$ from the messenger
scale $M$ down to the weak (or SUSY) scale $M_{SUSY}$. Whereas this RG
evolution sums up potentially large logarithms of the form
$\ln(M^2/M_{SUSY}^2)$, it does not describe contributions without such
logarithms which serve as boundary conditions for the RG evolution at
the scale $Q^2 = M^2$.

Note that both contributions (\ref{2.22}) and (\ref{2.23}) are
generated only at (or beyond) two loop order, and are hence suppressed
by additional factors $\lambda^2/(16\pi^2)^2 \times$ additional Yukawa
couplings. Compared to the effective SUSY breaking scale $m^2/(16 \pi^2
M)$, the contribution to the $A$ terms  (\ref{2.22}) is negligibly
small. However, the contribution (\ref{2.23}) to  $\Delta m_u^2 =
\Delta m_d^2$ can be of the same order as the two loop contributions
mediated by gauge interactions (see appendix A), if $\lambda$ is not
too small. Since the contribution (\ref{2.23}) to $\Delta m_u^2 =
\Delta m_d^2$ is typically negative, we will subsequently parametrize
it in terms of $\Delta_H$ defined as
\beq\label{2.24}
\Delta m_u^2 = \Delta m_d^2 = - \Delta_H \frac{\lambda^2}{(16\pi^2)^2}
M_{SUSY}^2
\eeq
with $M_{SUSY}=m^2/M$ as in appendix A, and $\Delta_H$ bounded from
above by $\Delta_H \lsim$ (Yukawa)$^2$ $\lsim {\cal{O}}(1)$.

To summarize this Section, within the class of models defined by the
superpotential (\ref{2.1}) one obtains in general, after integrating
out the messenger/sequestered sector, an effective NMSSM valid at
scales below the messenger scale $M$, which includes
\newline
a) the first two terms in the superpotential (\ref{2.1}),\newline
b) the soft SUSY breaking gaugino, squark, slepton and  Higgs
masses obtained by gauge mediation, which we recall for convenience in
appendix A,\newline
c) additional terms in the superpotential (\ref{2.14}) and
additional soft terms (\ref{2.15}),\newline
d) additional contributions to the soft SUSY breaking
Higgs masses as in (\ref{2.24}).

Note that neither an explicit $\mu$ term nor an explicit $m_3^2 \equiv
B\mu$ term are present at the messenger scale $M$. However, once the
above soft terms are used as boundary conditions for the RG evolution
from $M$ down to $M_{SUSY}$, a term of the form $m_3^2\,H_u H_d$ can be
radiatively generated in general. (In the appendix B, we recall the
$\beta$-functions of the parameters of the Higgs sector of the general
NMSSM. One finds that a non-vanishing parameter $m_S'^2$ generally
induces a non-vanishing $m_3^2$.)

Depending on the structure of the messenger/sequestered sector, many of
the terms in (\ref{2.16}) -- (\ref{2.21}) can be disallowed or
suppressed by discrete or approximate continuous symmetries. (Exact
continuous symmetries forbidding any of these terms would be
spontaneously broken in the physical vacuum, giving rise to an
unacceptable Goldstone boson.) An exception is the term (\ref{2.10})
leading to the soft singlet mass term (\ref{2.18}), which can never be
suppressed using symmetries. However, precisely this term is often
generated only to higher loop order and/or to higher order in an
expansion in $m/M$ as expected from na\"ive dimensional analysis
\cite{ue1995,dgs}. Finally we remark that terms of the form
$S F_S^* +h.c.$ (which give rise to the trilinear soft terms
(\ref{2.19})) will be suppressed if an $R$-symmetry is only weakly
broken in the scalar sector.

\section{Phenomenological analysis}

The purpose of this Section is the phenomenological analysis of various
scenarios within the class of models defined in Section 2, that differ
by the presence/absence of the different
terms (\ref{2.16}) to (\ref{2.21}) and (\ref{2.24}). 

To this end we employ a Fortran routine NMGMSB, that will be made
public on the NMSSMTools web page \cite{nmssmtools}. The routine NMGMSB
is a suitable generalization of the routine NMSPEC (available on the
same web site) towards the general NMSSM with soft SUSY breaking terms
specified by GMSB, i.e. it allows for a phenomenological analysis of
the class of models defined in Section 2. It requires the definition of
a model in terms of the parameter $\lambda$ and the soft SUSY breaking
and superpotential terms b) -- d) above. Since the coupling $\lambda$
at the effective SUSY breaking scale plays an important
phenomenological r\^ole (and in order to allow for comparisons with
other versions of the NMSSM as mSUGRA inspired), the coupling $\lambda$
on input is defined at an effective SUSY breaking scale $Q_{SUSY}$
given essentially by the squark masses. The remaining input parameters,
notably the soft SUSY breaking terms listed in appendix A and in
(\ref{2.16}) -- (\ref{2.21}), are defined at a unique messenger scale
$M$.

The RG equations are then integrated numerically from $M$ down to
$Q_{SUSY}$. Additional input parameters are, of course, $M_Z$, and also
$\tan\beta$ (at the scale $M_Z$). Similar to the procedure employed in
NMSPEC, the minimization equations of the effective Higgs potential --
including radiative corrections as in \cite{nmssmtools} -- can then be
solved for the Yukawa coupling $\kappa$ in the superpotential
(\ref{2.1}), and for the SUSY breaking singlet mass $m_S^2$
(\ref{2.18}) or, if $m_S^2$ is fixed as input, for $\xi_S$.  (If
specific values for $\kappa$, $m_S^2$ and $\xi_S$ at the scale $M$
are desired as input, this procedure is somewhat inconvenient. Then,
one would have to scan over at least some of the other input parameters  and
select points in parameter space where $\kappa$, $m_S^2$ or $\xi_S$ --
which are given at the scale $M$ as output -- are close enough to the
desired numerical values.) Since the gauge and SM Yukawa couplings are
defined at the scale $M_Z$, a few iterations are required until the
desired boundary conditions at $M_Z$ and $M$ are  simultaneously
satisfied.

After checking theoretical constraints as the absence of deeper minima
of the effective potential and Landau singularities below $M$, the
routine proceeds with the evaluation of the physical Higgs masses and
couplings (in\-cluding radiative corrections as in  \cite{nmssmtools})
and the sparticle spectrum including pole mass corrections. Then,
phenomenological constraints can be checked:
\newline
-- Higgs masses, couplings and branching ratios are compared to
constraints from LEP, including constraints on unconventional Higgs
decay modes \cite{lhg} relevant for the NMSSM; \newline
-- constraints from $B$-physics are applied as in \cite{bphys}, and the
muon anomalous magnetic moment is computed.

Subsequently we investigate several scenarios, for which many (but
different) terms in the list (\ref{2.16}) -- (\ref{2.21}) vanish or
are negligibly small.

\subsection{Scenarios with tadpole terms}

The tadpole terms $\xi_F$ in $\Delta W$ in (\ref{2.14}) and $\xi_S$ in
$\Delta V_{soft}$ in (\ref{2.15}) will trigger a nonvanishing VEV of
$S$. However, as it becomes clear from (\ref{2.17}) and (\ref{2.21}),
these tadpole terms -- if not forbidden by symmetries -- tend to be too
large: the scale of the soft SUSY breaking gaugino, squark, slepton and
Higgs masses in GMSB models is given by $M_{SUSY} \sim m^2/M$ (together
with an additional loop factor $(16\pi^2)^{-1}$, see appendix A).
Written in terms of $M$ and $M_{SUSY}$, the maximally possible order of
magnitude of the supersymmetric and soft SUSY breaking tadpole terms
are $\xi_F \sim m^2 \sim M M_{SUSY}$ and  $\xi_S \sim m^4/M \sim M
M_{SUSY}^2$. If $M \gg M_{SUSY}$, which will generally be the case,
these tend to be larger than the desired orders of
magnitude $\xi_F \sim M_{SUSY}^2$ and $\xi_S \sim M_{SUSY}^3$.
(This problem is similar to the $\mu$ and $B \mu$ problem in
the MSSM with GMSB, see \cite{dgp}.) Hence one has to assume that
these terms are suppressed, e.g. generated to higher loop order only as
in \cite{des}, or involve small Yukawa couplings. Let us study the
latter scenario quantitatively in a simple model \cite{ue1995}: let us
assume that the singlet superfield couples directly to $n_5$ pairs of
messengers $\widehat{\phi}$, $\widehat{\overline{\phi}}$ (in
$\underline{5}$ and $\overline{\underline{5}}$ representations under
$SU(5)$) due to a term
\beq\label{3.1}
-\eta\widehat{S}\widehat{\overline{\phi}}\widehat{\phi}
\eeq
in the superpotential $\widetilde{W}$ in (\ref{2.1}). Then, one loop
diagrams generate \cite{ue1995}
\beq\label{3.2}
\xi_F = n_5\frac{\eta}{8\pi^2} m^2 \ln\left(\Lambda^2/M^2\right)
%\xi_F = \frac{\eta}{16\pi^2} m^2 \ln\left(\Lambda^2/M^2\right)
\eeq
and
\beq\label{3.3}
\xi_S = -n_5\frac{\eta}{16 \pi^2} \frac{m^4}{M}
%\xi_S = \frac{\eta}{16 \pi^2} \frac{m^4}{M}
\eeq
in agreement with the power counting rules (\ref{2.17}) and
(\ref{2.21}). (The UV cutoff $\Lambda$ appears in (\ref{3.2}) only if
the SUSY breaking $F_X$ is ``hard'' in the sense discussed in Section
2; otherwise the logarithm in (\ref{3.2}) should be replaced by a
number of ${\cal O}(1)$.)

Below, we consider a mass splitting $m^2 \sim 8\times 10^{10}$ GeV$^2$
among the messenger scalars and pseudoscalars, and a messenger scale $M
\sim 10^6$ GeV. Then, for $\ln\left(\Lambda^2/M^2\right) \sim 3$, a
Yukawa coupling $\eta \sim 2\times 10^{-6}$ generates $\xi_F \sim (150\
\mathrm{GeV})^2$ and $\xi_S \sim -(1\ \mathrm{TeV})^3$. We find that
these orders of magnitude for $\xi_F$ and $\xi_S$ are perfectly
consistent with a phenomenologically viable Higgs sector. Given the
presence of small Yukawa couplings in the Standard Model, and the
possibility of obtaining additional symmetries in the limit of
vanishing $\eta$, we do not consider $\eta \sim 10^{-6} - 10^{-5}$ as
particularly unnatural.

The coupling (\ref{3.1}) also gives rise to a positive SUSY breaking 
mass squared
\beq\label{3.4}
%m_S^2 = \frac{\eta^2}{16\pi^2} \frac{m^6}{M^4}
m_S^2 = n_5\frac{\eta^2}{4\pi^2} \frac{m^6}{M^4}
\eeq
for the singlet $S$. Under the assumption of such small values for
$\eta$, this term  is numerically negligible
(as well as contributions to $A_\lambda, A_\kappa,
\mu', m_S'^2, \Delta_H$ and two loop contributions to $m_S^2$ of ${\cal
O}(m^4/M^2)$).

Hence in the following we will concentrate on models where, among the
terms in (\ref{2.16}) -- (\ref{2.21}) and (\ref{2.24}), only $\xi_F$
and $\xi_S$ are nonvanishing. (These models are then similar to the
ones denoted as ``nMSSM'' in \cite{nnmssm}. However, given the present
constraints on the soft terms we found that a term $\sim \kappa$ in the
superpotential (\ref{2.1}) is required for the stability of the scalar
potential.) The remaining free parameters are $\tan\beta$, $\lambda$,
$M$, $m^2/M$ and $\xi_F$: since $m_S^2$ is fixed as input at the scale
$M$ (where $m_S^2=0$), the equation following from the minimization of
the potential w.r.t. $S$ can be used to determine~$\xi_S$.

Quite generally, there exist two distinct allowed regions in the
parameter space, which differ how the lightest scalar Higgs
mass $m_{h_1}$ satisfies the LEP bound of $\sim 114$~GeV: \newline
a) region A at low $\tan\beta$ and large $\lambda$, where the NMSSM
specific contributions to the lightest Higgs mass allow for values
above above 114 GeV. Low values of $\tan\beta$ demand that the
messenger scale $M$ is not too large: $\tan\beta \sim 1$ requires
$m_u^2 \sim m_d^2$ at the SUSY scale, but the RG equation for $m_u^2$
differs from the one for $m_d^2$ by the presence of the top Yukawa
coupling (which is particularly large for small $\tan\beta$). Thus the
range of the RG running should not be too big, i.e. the scale $M$ should
not be too far above the SUSY scale.\newline
b) region B at large $\tan\beta$, where the messenger scale $M$ is
quite large (typically $\sim 10^{13}$ GeV) resulting in stop masses in
the 1.5--2 TeV range. Then the top/stop radiative corrections to the
lightest Higgs mass can lift it above 114 GeV without the need for
NMSSM specific contributions. (At large $\tan\beta$, $\lambda$ does not
increase the lightest Higgs mass; on the contrary, large values of
$\lambda$ lower its mass through an induced mixing with the
singlet-like scalar. Hence, $\lambda$ must be relatively small here.)
However, in the present context one finds from (\ref{3.2}) and
(\ref{3.3}) that such large values for $M$ (with fixed $m^2/M \sim
10^5$~GeV) would require extremely small values for $\eta$. For this
reason we confine ourselves to region A in the following.

In region A, the LEP bound on $m_{h_1}$ requires $\tan\beta$  to be
smaller than $\sim 2$, and $\lambda$ larger than $\sim 0.45$.
Subsequently we investigate the interval $0.45 < \lambda < 0.6$ and
$\tan\beta > 1.2$, where perturbativity in the running Yukawa couplings
$\lambda$, $\kappa$ and $h_t$ is guaranteed at least up to the
messenger scale $M$. If we na\"ively extrapolate the RGEs beyond the
scale $M$ (taking the contributions of the messenger fields to the
running gauge couplings into account), perturbativity in the running
Yukawa couplings is usually not satisfied up to the GUT scale in region
A (in contrast to scenarios where $M \sim 10^{13}$ GeV). There exist
different possible solutions to this problem: first, additional matter
could be present at the messenger scale, charged under the SM gauge
groups. Then, SM gauge couplings can become large (at the boundary of
perturbativity) below the GUT scale, and since they  induce a negative
contribution to the $\beta$~functions for $h_t$ and $\lambda$, they
could help to avoid a Landau singularity in the Yukawa sector below
$M_{GUT}$. Another attitude would be to assume that a strongly
interacting sector (possibly responsible for the breaking of
supersymmetry) exists at or above the messenger scale $M$; then the
singlet $S$, for example, could turn out to be a composite state which would
imply a compositeness condition equivalent to Landau singularities in
the Yukawa couplings of $S$ at the corresponding scale (without
affecting, at the one loop level, the grand unification of the SM gauge
couplings).

Within the region $1.2 < \tan\beta < 2$ and $0.45 < \lambda < 0.6$, a
wide range of the remaining parameters $M$, $m^2/M$ and $\xi_F$
satisfies all phenomenological constraints. Subsequently we fix these
parameters near the center of the allowed range:  $M=10^6$ GeV,
$m^2/M=8\times 10^4$ GeV and $\xi_F=3\times 10^4$ GeV$^2$, and vary
$\tan\beta$ and $\lambda$ in the above intervals (taking, for
simplicity, $n_5 = 1$).

The allowed range of $\tan\beta$ ($\tan\beta < 1.6$ for these values
for $M$, $m^2/M$ and $\xi_F$) and $\lambda$ (actually $\lambda \gsim
0.5$) is shown in Fig. 2; the upper limit on $\tan\beta$ originates
from the LEP bound on the lightest Higgs mass $m_{h_1}$. This becomes
evident from Fig. 3, where we show the range of $m_{h_1}$ (for various
values of $\lambda$, larger values of $\lambda$ corresponding to larger
values of $m_{h_1}$) as a function of $\tan\beta$. If we would allow for
larger values of $\lambda$ (and/or smaller values of $\tan\beta$),
larger values for  $m_{h_1}$ would be possible.

\begin{figure}[ht]
%\vskip 1.0cm
\begin{center}
\includegraphics[angle=-90,scale=0.5]{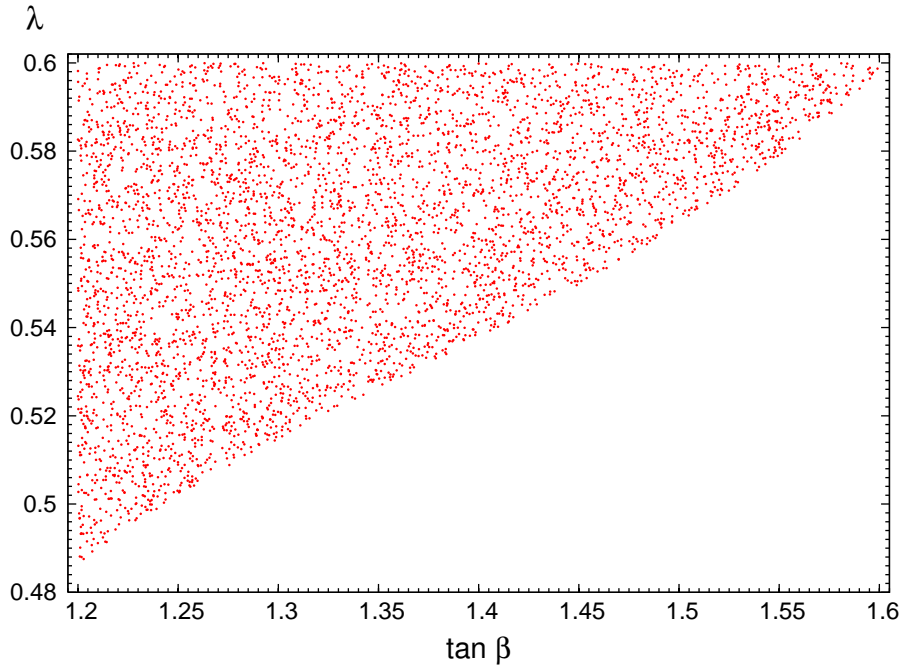}
\caption{Allowed values of $\lambda$ as a function of $\tan\beta$ for 
$M=10^6$ GeV, $M_{SUSY}=m^2/M=8\times 10^4$ GeV and $\xi_F=3\times
10^4$ GeV$^2$.}
\label{1.2f}
\end{center}
%\vspace*{5mm}
\end{figure}

\begin{figure}[ht]
%\vskip 1.0cm
\begin{center}
\includegraphics[angle=-90,scale=0.5]{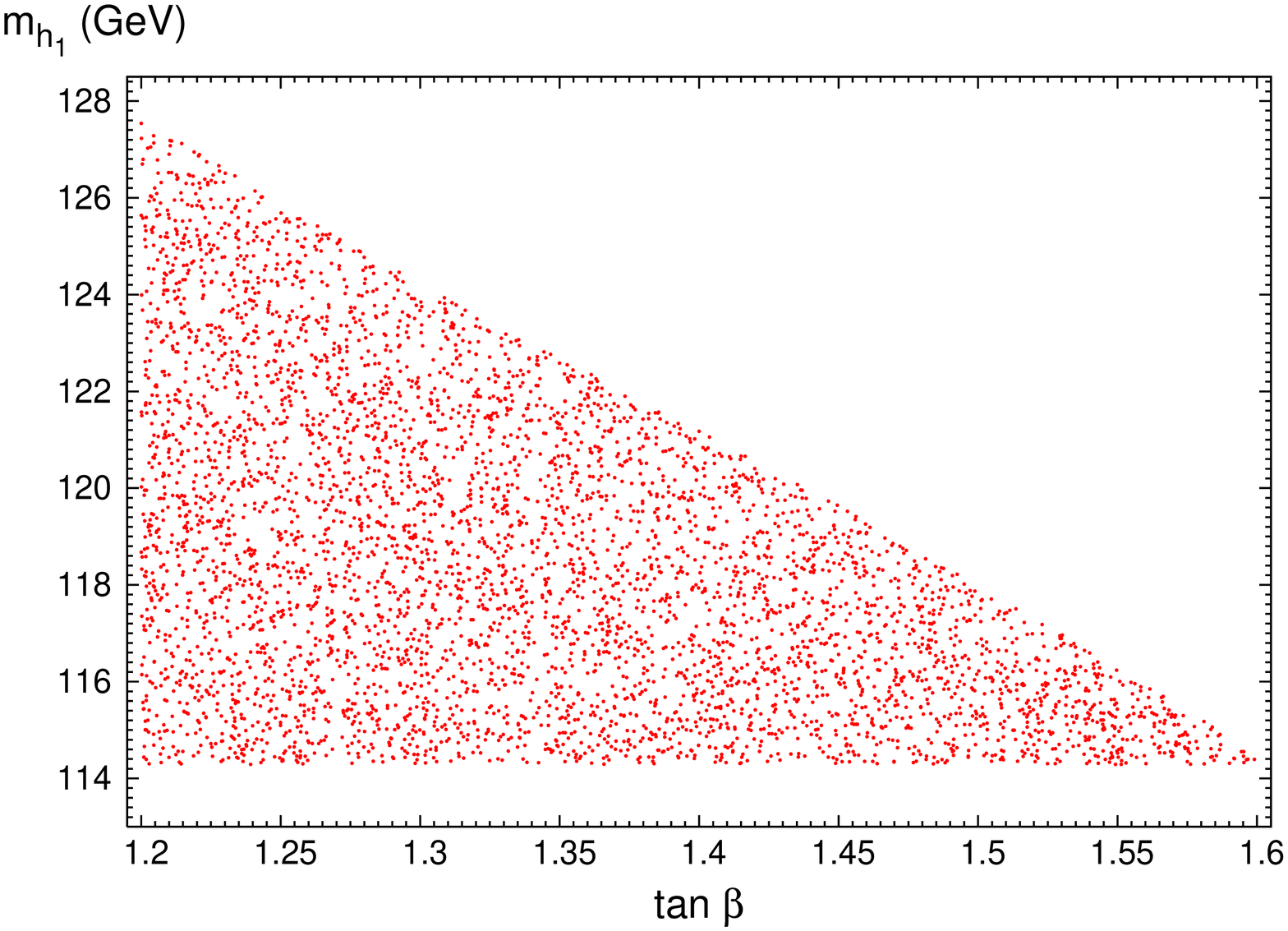}
\caption{The lightest Higgs mass as a function of $\tan\beta$ for the
same parameters as in Fig.~2, larger values of $m_{h_1}$ corresponding
to larger values of $\lambda$.}
\label{1.3f}
\end{center}
%\vspace*{5mm}
\end{figure}

In Fig. 4 we display the charged Higgs mass $m_{h^\pm}$ (practically
degenerated with a scalar with mass $m_{h_2}$ and a pseudoscalar with
mass $m_{a_2}$), the singlet-like scalar mass $m_{h_3}$ and the
singlet-like pseudoscalar mass $m_{a_1}$, all of which are nearly
independent of $\lambda$. For small $\tan\beta$ the large values of the
Higgs masses indicate that this region is implicitly more fine tuned. 
The remaining sparticle spectrum is essentially specified by $M$ and
$m^2/M$, and hardly sensitive to $\tan\beta$ and $\lambda$ within the
above intervals: 
\bea
\mathrm{Bino:} && \sim 105\ \mathrm{GeV}\nn\\
\mathrm{Winos:} && \sim 200\ \mathrm{GeV}\nn\\
\mathrm{Higgsinos:} && \sim 670 - 1000\ \mathrm{GeV}\nn\\
\mathrm{Singlino:} && \sim 900 - 1800\ \mathrm{GeV}\nn\\
\mathrm{Sleptons:} && \sim 140 - 290\ \mathrm{GeV}\nn\\
\mathrm{Squarks:} && \sim 640 - 890\ \mathrm{GeV}\nn\\
\mathrm{Gluino:} && \sim 660\ \mathrm{GeV}\nn
\eea
(Due to the small value of $\tan\beta$ in this scenario, the
supersymmetric contribution to the muon anomalous magnetic moment is
actually too small to account for the presently observed
deviation w.r.t. the Standard Model.)

\begin{figure}[ht]
%\vskip 1.0cm
\begin{center}
\includegraphics[angle=-90,scale=0.5]{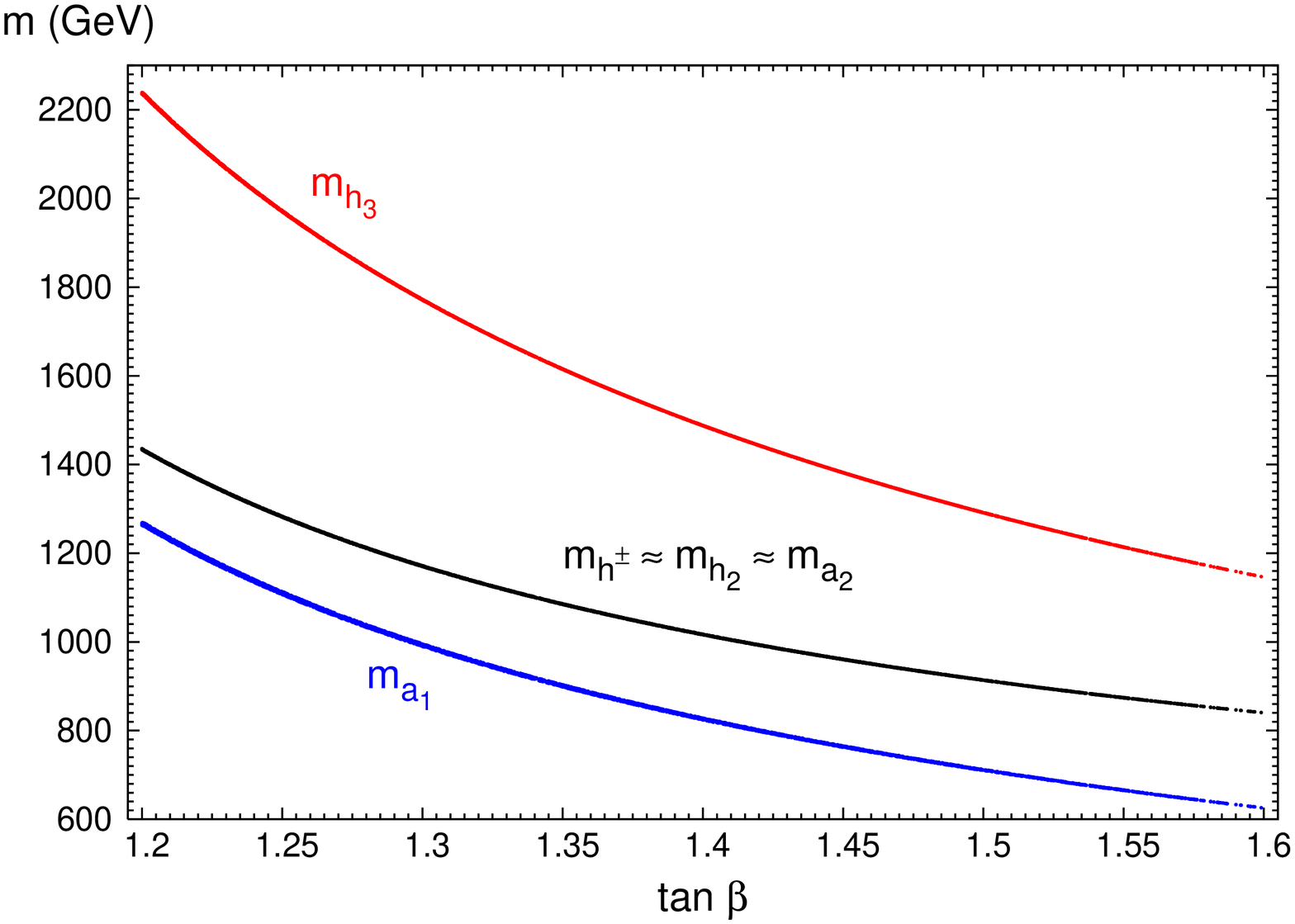}
\caption{Heavy Higgs masses as a function of $\tan\beta$ for the
same parameters as in Fig.~2.}
\label{1.4f}
\end{center}
%\vspace*{5mm}
\end{figure}

In Fig. 5, we give the values of $\xi_S$ (at the scale $M$), which are
obtained as an output as function of $\tan\beta$. Within the model
corresponding to (\ref{3.1}) -- (\ref{3.3}) above, one can easily
deduce the Yukawa coupling $\eta$ from $\xi_S$ using (\ref{3.3})
resulting in $\eta$ varying in the range $2\times 10^{-6}$ (for
$\tan\beta = 1.6$) to $10^{-5}$ (for $\tan\beta = 1.2$). The
corresponding value of $\ln\left(\Lambda^2/M^2\right)$ can then be
deduced from (\ref{3.2}), with the conclusion that 
$\ln\left(\Lambda^2/M^2\right)$ should assume values in the range 1 to
4 -- a reasonable result, by no means guaranteed, that we consider as a
strong argument in favour of such a simple model.

\begin{figure}[ht]
%\vskip 1.0cm
\begin{center}
\includegraphics[angle=-90,scale=0.5]{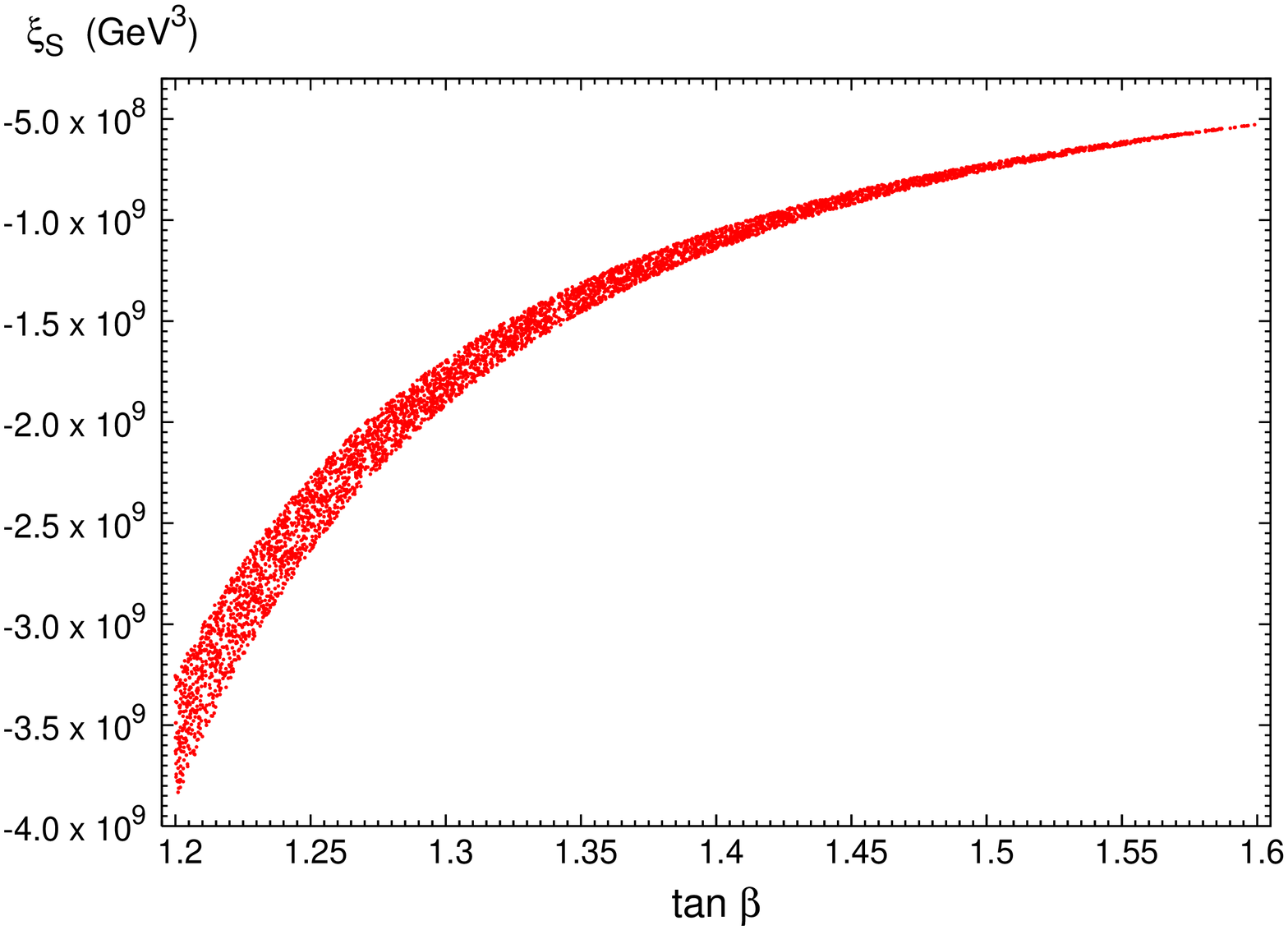}
\caption{$\xi_S$ as a function of $\tan\beta$ for the same parameters
as in Fig. 2.}
\label{1.5f}
\end{center}
%\vspace*{5mm}
\end{figure}

Finally we note that for larger values of $n_5$ (as $n_5=3$), $M$ (as
$M = 2\times 10^{10}$) and $\xi_F$ (as $\xi_F= 10^5$ GeV$^2$, see also
the next subsection) phenomenologically viable regions in parameter
space exist where the running Yukawa couplings $\lambda$, $\kappa$ and
$h_t$ remain perturbative up to $M_{GUT}$. Within the model above,
these scenarios would require an even smaller Yukawa coupling $\eta$, 
$\eta \sim 10^{-8}$.

\subsection{Scenarios without tadpole terms}

Scenarios without tadpole terms have been proposed in \cite{gr}. If
the number of messengers is doubled ($n_5 = 2$), i.e. introducing
$\widehat{\Phi}_1$, $\widehat{\overline{\Phi}}_1$,  $\widehat{\Phi}_2$
and $\widehat{\overline{\Phi}}_1$, these can couple to $\widehat{S}$
and to the spurion $\widehat{X}$ in such a way that a discrete $Z_3$
symmetry is left unbroken by the VEV of $\widehat{X}$ \cite{gr}:
\beq\label{3.5}
\widetilde{W}=\widehat{X}\left(\widehat{\overline{\Phi}}_1
\widehat{\Phi}_1 + \widehat{\overline{\Phi}}_2 \widehat{\Phi}_2 \right)
+\eta\widehat{S}\,\widehat{\overline{\Phi}}_1\widehat{\Phi}_2
\eeq

Then tadpole terms $\sim \xi_F$ and $\sim \xi_S$ are disallowed, and
the Yukawa coupling $\eta$ can be much larger. These scenarios have
been recently investigated in \cite{dgs} (see also \cite{gkr}), where
the $SU(5)$ breaking (generated via the RG equations between $M_{GUT}$
and $M$) inside $\eta\widehat{S}\,\widehat{\overline{\Phi}}_1
\widehat{\Phi}_2$ has been taken into account.

For larger values of $\eta$, messenger loops generate non-negligible
values for the singlet mass $m_S^2$ (\ref{2.18}), trilinear $A$-terms
(\ref{2.19}) and corrections $\Delta m_u^2 = \Delta m_d^2$ as in
(\ref{2.24}) at the scale $M$ \cite{dgs}. Phenomenologically viable
regions in parameter space have been found in \cite{dgs}, where the
parameters $M$ and $M_{SUSY}$ have been chosen as $M = 10^{13}$~GeV and
$M_{SUSY}=m^2/M = 1.72 \times 10^{5}$~GeV. The stop masses are quite
large (up to $\sim 2$~TeV) such that the stop/top induced radiative
corrections to $m_{h_1}$ lift it above the LEP bound of $\sim 114$~GeV.

We have re-investigated this scenario in a somewhat simpler setup:
first we observe that the generated values for $A_\kappa$ and
$\Delta_H$, in the notation (\ref{2.19}) and (\ref{2.24}), are always
related by
\beq\label{3.6}
A_\kappa = -\frac{3}{16 \pi^2} \Delta_H M_{SUSY}
\eeq
(with $\Delta_H = 2\xi_D^2 + 3\xi_T^2$ in the notation of \cite{dgs},
where $\xi_{D,T}$ denote Yukawa couplings corresponding to our $\eta$
in (\ref{3.5}). At $M_{GUT}$ one has $\xi_D = \xi_T \equiv \xi_U$
\cite{dgs}.) The singlet mass at the scale $M$ is then of the order
\beq\label{3.7}
m_S^2 \simeq \frac{1}{(16 \pi^2)^2} \left(\frac{7}{5} \Delta_H^2 -
\frac{1}{5} (16 g_3^2 + 6g_2^2 +\frac{10}{3}g_1^2) \Delta_H -4\kappa^2
\Delta_H \right) M_{SUSY}^2\ ,
\eeq
where we have neglected the $SU(5)$ breaking among the Yukawas at the
scale $M$.

We tried to reproduce the three phenomenologically viable regions in
parameter space studied in \cite{dgs}: region I where $\xi_U \ll 1$,
region III where $0.6 \lsim \xi_U \lsim 1.1$, and region II where $1.3
\lsim \xi_U \lsim 2$. We observe, however, that for $\xi_U \gsim 0.7$
(or $\Delta_H \gsim 1.5$ after taking the running of $\xi_U$ between
$M_{GUT}$ and $M$ into account) the generated value for
$\left|A_\kappa\right|$ from (\ref{3.6}) exceeds $\sim 5$ TeV at $M$
(still $\gsim 2$ TeV at the weak scale), which we interpret as a
certain amount of fine tuning between the remaining parameters of
the Higgs potential. We will not consider the region II below.
Note that, as in \cite{dgs}, we obtain $\kappa$ as an output (from the
minimization equations of the Higgs potential with $M_Z$ as input),
which can hide the fine tuning required.

Limiting ourselves to $\Delta_H \lsim 1.5$ ($\left|A_\kappa\right|
\lsim 5$ TeV), we were able to confirm the region~I. In Table~1 we show
the Higgs spectrum, and in Table~2 the essential features of the
corresponding sparticle spectrum for a representative point P1 in
region~I, where  $A_\kappa = -160$~GeV, $\Delta_H = 0.1$, $\lambda =
0.02$ and $\tan\beta = 6.6$ (leading to $m_S^2 \sim -2.8 \times
10^5$~GeV$^2$ in agreement with (\ref{3.7})). The point P2 in Tables~1
and 2 is in the region III of \cite{dgs}: there one has $A_\kappa =
-4.77$~TeV, $\Delta_H = 1.46$,  $\lambda = 0.5$ and $\tan\beta = 1.64$
($m_S^2 \sim -5.3 \times 10^6$~GeV$^2$). We see that, in spite of stop
masses in the 2 TeV region, $m_{h_1}$ is not far above the LEP bound.
On the other hand these results confirm the phenomenological viability
of the scenario proposed in \cite{gr,dgs}. (However,  due to the very
heavy sparticle spectrum the supersymmetric contribution to the muon
anomalous magnetic moment is still too small to account for the presently
observed deviation w.r.t. the Standard Model.)

\begin{table}[!ht]
\caption{Input parameters and Higgs masses for five specific points.}
\vspace*{-5mm}
\label{table1}
\vspace{3mm}
\footnotesize
%\vspace*{-22mm}\hspace*{-15mm}
\begin{center}
\begin{tabular}{|l|r|r|r|r|r|}
\hline
{\bf Point} & P1 & P2 &  P3 & P4 & P5
\\\hline
{\bf Input parameters }
\\\hline
Messenger scale $M$ (GeV) & $10^{13}$ & $10^{13}$ & $4\times 10^{8}$ 
& $3\times 10^{7}$  & $5\times 10^{14}$ 
\\\hline
$M_{SUSY} = m^2/M$ (GeV)  & $1.72\times 10^{5}$ & $1.72\times 10^{5}$ 
&  $3.2\times 10^{4}$ & $3.5\times 10^{4}$ & $7.5\times 10^{4}$
\\\hline
$\tan\beta$  & 6.6 & 1.64 & 1.6 & 1.9 & 40
\\\hline
$n_5$ & 2 & 2 & 2 & 2 & 2
\\\hline
$\lambda$ & 0.02 & 0.5 & 0.6 & 0.6 & 0.01
\\\hline
$A_\kappa$ (GeV) & -160 &-4770 & 0 & 0 & 0
\\\hline
$\Delta_H$ & 0.1 & 1.46 & 0 & 0 & 0
\\\hline 
$m_S^2$ (GeV$^2$) & $-2.8\times  10^5$ & $-5.3\times  10^6$ 
& $-4.3\times  10^5$ & $-2.1\times 10^{5}$ & $-5.0\times 10^{3}$ 
\\\hline\hline
{\bf CP even Higgs masses}
\\\hline
$m_{h^0_1}$ (GeV) & 116.1 & 115.8 & 115.5 & 96.1 &94.5
\\\hline
$m_{h^0_2}$ (GeV) & 794 & 2830 & 607 & 514 & 120
\\\hline
$m_{h^0_3}$ (GeV) & 1762 & 3411 & 717 & 579 & 603
\\\hline\hline
{\bf CP odd Higgs masses}
\\\hline
$m_{a^0_1}$ (GeV) & 448 & 2842 & 40.5 & 11.5 & 1.1
\\\hline
$m_{a^0_2}$ (GeV) & 1761& 3662  & 628 & 546 & 603
\\\hline\hline
{\bf Charged Higgs mass} 
\\\hline\hline
$m_{h^\pm}$ (GeV) & 1764 & 2862 & 619 & 535 & 613
\\\hline
\end{tabular}\end{center}
\end{table}

\begin{table}[!ht]
\caption{Some sparticle masses and components for the five specific 
points of Table~1. The chargino masses are close to the
wino/higgsino-like neutralino masses, the right-handed/left-handed 
slepton masses close
to the stau$_1$/stau$_2$ masses, and the remaining squark masses are of
the order of the gluino mass.}
\vspace*{-5mm}
\label{table2}
\vspace{3mm}
\footnotesize
%\vspace*{-22mm}\hspace*{-15mm}
\begin{center}
\begin{tabular}{|l|r|r|r|r|r|}
\hline
{\bf Point} & P1 & P2 &  P3 & P4 & P5
\\\hline
{\bf Neutralinos }
\\\hline
$\chi_1$ mass (GeV) & 467 & $469$ & $80.5$ 
& $88.3$  & $101$ 
\\\hline
Dominant component  & bino & bino
& bino  & bino & singlino
\\\hline
$\chi_2$ mass (GeV) & 839 & $890$ & $152$ 
& $166$  & $200$
\\\hline
Dominant component  & singlino & wino
& wino  & wino & bino
\\\hline
$\chi_3$ mass (GeV) & 882 & $2322$ & $463$ 
& $428$  & $380$ 
\\\hline
Dominant component  & wino & higgsino
& higgsino & higgsino & wino
\\\hline
$\chi_4$ mass (GeV) & 1432& $2325$ & $476$ 
& $438$  & $675$ 
\\\hline
Dominant component  & higgsino &   higgsino
& higgsino  & higgsino & higgsino
\\\hline
$\chi_5$ mass (GeV) & 1440 & $4019$ & $721$ 
& $572$  & $685$ 
\\\hline
Dominant component  & higgsino & singlino 
& singlino  & singlino & higgsino
\\\hline\hline
Stau$_1$ mass (GeV) & 692 & $693$ & $100$ 
& $103$  & $260$ 
\\\hline
Stau$_2$ mass (GeV) & 1100 & $1096$ & $188$ 
& $198$  & $514$ 
\\\hline
Stop$_1$ mass (GeV) & 1931 & $1819$ & $376$ 
& $459$  & $872$ 
\\\hline
Gluino mass (GeV) & 2389 & $2386$ & $522$ 
& $569$  & $1117$ 
\\\hline
\end{tabular}\end{center}
\end{table}

\subsection{Scenarios without tadpole and $A$-terms}

The scenario discussed in the previous subsection belongs to those
where many (actually most) of the operators (\ref{2.7}) -- (\ref{2.12})
and (\ref{2.22}) -- (\ref{2.23}) are forbidden by a discrete $Z_N$
symmetry, which is left unbroken in the messenger/seques\-tered sector,
but under which $\widehat{S}$ carries a non-vanishing charge. In the
above case -- where $Z_N$ is {\it not} an $R$-symmetry -- all soft terms
$m_S^2$, $A_\kappa = 3 A_\lambda$ and the parameter $\Delta_H$ in
(\ref{2.24}) will in general be non-vanishing (all others being
forbidden).

The fate of $R$-symmetries in the context of gauge mediation has recently
been reviewed in \cite{adjk}. In the case of spontaneous 
breaking within the messenger/sequestered sector \cite{adjk1},
$R$-symmetry violating terms in the effective low energy theory will be
suppressed relative to $R$-symmetry conserving terms. Then, the trilinear
terms  $A_\kappa = 3 A_\lambda$ (\ref{2.19}) will be  negligibly small.
Although the $R$-symmetry breaking gaugino masses will typically also be
smaller than the scalar masses at the messenger scale \cite{adjk}, we
will consider in this subsection an illustrative scenario which is just
a limiting case of the one previously discussed.  

We will investigate the case where the trilinear terms
vanish, and where only $m_S^2$  (which can never be forbidden by
symmetries) assumes natural values at the messenger scale $M$. For
simplicity, we will allow for standard gaugino masses (and the usual
scalar masses) as given in appendix A.
Now, the scalar sector of the NMSSM has an exact $R$-symmetry at the
scale $M$, with identical charges for all superfields. Given that
gaugino masses break this $R$-symmetry, radiative corrections (the RG
running between $M$ and the weak scale) induce $R$-symmetry violating
trilinear terms in the scalar sector. If $M$ is not too large or if
$\lambda$, $\kappa$ are small, these trilinear terms remain
numerically small, and the $R$-symmetry in the scalar sector is only
weakly broken. Given that this approximate $R$-symmetry is spontaneously
broken at the weak scale by the VEVs of $H_u$, $H_d$ and $S$, a pseudo
Goldstone boson (a pseudo $R$-axion \cite{raxion}) appears in the
spectrum. Light pseudoscalars can lead to a reduction of the LEP
constraints on $m_{h_1}$, and have recently been the subject of various
investigations \cite{lighthiggs}.

In what follows we study the phenomenological viability of such scenarios,
which are defined by having all terms (\ref{2.16}) -- (\ref{2.21})
vanish except for $m_S^2$ (but vanishing $A_\kappa$, $A_\lambda$). For
simplicity we will also assume that $\Delta_H$ in (\ref{2.24}) is
negligibly small. Then, the model is completely specified by $\lambda$,
$\tan\beta$ and the scales $M$ and $M_{SUSY}$ (recall that $\kappa$ and
$m_S^2$ can be obtained from the minimization equations in terms of
$M_Z$ and of the other parameters). Again we found that two completely
different regions in parameter space are phenomenologically viable.

As before, the first region is characterized by small values of
$\tan\beta$ ($\tan\beta \lsim 2$) and large values of $\lambda$.
Relatively large negative values for the soft mass $m_S^2$ for the
singlet of the order $m_S^2 \sim -(600\ \mathrm{GeV})^2$ are required
at the scale $M$ in order to generate the required VEV of~$S$. The mass
$m_{a_1}$ of the lightest CP-odd scalar varies in the range $0 <
m_{a_1} < 50$~GeV, where the larger values are obtained for larger
messenger scales $M \sim 10^9$~GeV: then the RG evolution generates
relatively large values $A_\lambda \sim 25$ GeV at the weak scale
(whereas $A_\kappa$ remains very small), and this breaking of the
$R$-symmetry induces a relatively large mass for the pseudo $R$-axion. On
the other hand, arbitrarily small values for $A_\lambda$ and hence for
$m_{a_1}$ can be obtained without any fine tuning for lower messenger
scales $M$. In all cases we find that the lightest CP-even (SM like)
scalar $h_1$ dominantly decays (with branching ratios of $\sim$ 80\%)
into $h_1 \to a_1 a_1$, which allows for $m_{h_1} < 114$~GeV consistent
with LEP constraints.

For given $\lambda$, $m_{h_1}$ is nearly independent of the scales $M$
and $M_{SUSY}$, but decreases with $\tan\beta$. In Fig.~6 we show a
scatterplot for $m_{h_1}$ as a function of $\tan\beta$, which is
obtained for $\lambda = 0.6$, varying $M$ in the range $10^7\
\mathrm{GeV} < M < 5\times 10^9\ \mathrm{GeV}$ and $M_{SUSY}$ in the
range $3.3 \times 10^4\ \mathrm{GeV} < M_{SUSY} < 4.3\times 10^4\
\mathrm{GeV}$. All points displayed satisfy LEP and $B$-physics
constraints. (We have chosen $n_5 = 2$ messenger multiplets, but
similar results can be obtained -- for slightly different ranges of $M$
and $M_{SUSY}$ -- for $n_5 = 1$.)

\begin{figure}[ht]
%\vskip 1.0cm
\begin{center}
\includegraphics[angle=-90,scale=0.5]{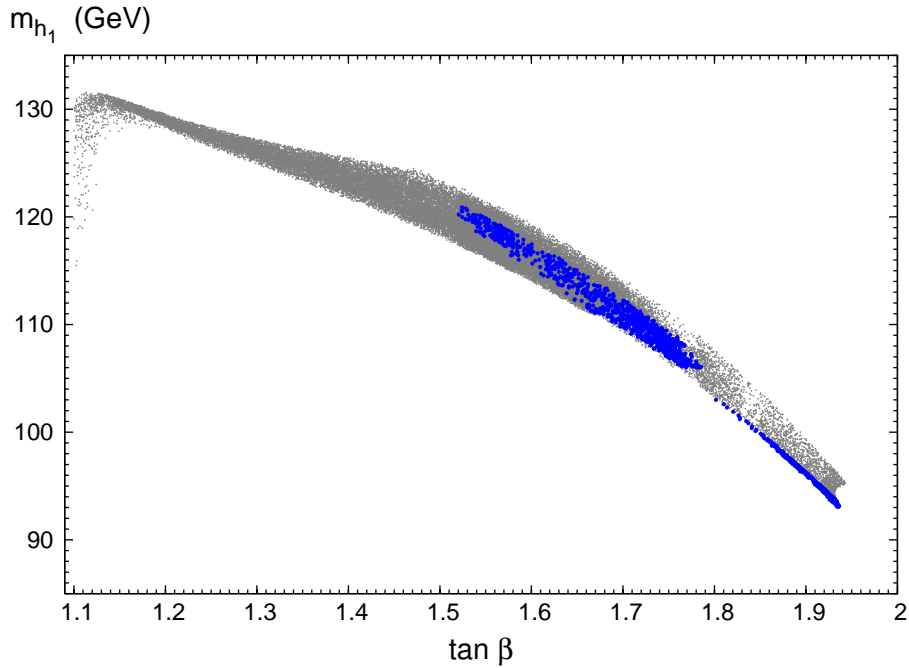}
\caption{$m_{h_1}$ as a function of $\tan\beta$ for $\lambda = 0.6$,
$10^7\ \mathrm{GeV} < M < 5\times 10^9\ \mathrm{GeV}$ and $3.3 \times
10^4\ \mathrm{GeV} < M_{SUSY} < 4.3\times 10^4\ \mathrm{GeV}$.
Points where, in addition to all 
LEP and $B$-physics constraints, the SUSY contribution to the muon 
anomalous magnetic moment can (fails to) account for the presently observed 
deviation with respect to the Standard Model are denoted in blue/darker 
(gray/lighter) color.}
\label{1.6f}
\end{center}
%\vspace*{5mm}
\end{figure}

%\vskip -5mm

In the region $\tan\beta \gsim 1.7$ (where $m_{h_1} \lsim 108$~GeV) LEP
constraints are satisfied only for $m_{a_1} \lsim 11$~GeV, so that $a_1
\to bb$ decays are forbidden and the dominant decays of $h_1$ are $h_1
\to a_1 a_1 \to 4\ \tau$ (still requiring $m_{h_1} \gsim
88$~GeV \cite{lhg}). For $\tan\beta \lsim 1.7$, the dominant decays of
$h_1$ are $h_1 \to a_1 a_1 \to 4\ b$, in which case LEP constraints
allow for $m_{h_1}$ as low as $\sim 108$~GeV. The complete
theoretically possible range for $m_{a_1}$ is now allowed by LEP.
(Fixing, e.g., $M = 10^8\ \mathrm{GeV}$, the complete range $1.2 \lsim
\tan\beta \lsim 1.7$ is compatible with LEP constraints on the Higgs
sector within the above range of $M_{SUSY}$. For smaller $\tan\beta$,
however, the hidden fine tuning becomes quite large.)

Now, in some regions in parameter space, the supersymmetric
contribution to the muon anomalous magnetic moment is $\gsim 10^{-9}$,
which accounts for the presently observed deviation with respect to
the Standard Model. 
The blue (darker) points in Fig.~6 (which appear only for
$\tan\beta \gsim 1.5$) satisfy this condition. In Tables~1 and 2 we
present the Higgs and sparticle spectrum for points P3 (with $\tan\beta =
1.6$) and P4 (with $\tan\beta = 1.9$), which are inside the blue region
of Fig.~6.

Another interesting region in parameter space is characterized by large
values of $\tan\beta$ ($\tan\beta \gsim30$) and small values of
$\lambda$ ($\lambda \sim 10^{-2}$), associated with small values of
$\kappa$ ($\kappa \lsim 10^{-3}$). 
In this case, comparatively small negative values for
the soft mass $m_S^2$ for the singlet of the order $m_S^2 \sim -(70\
\mathrm{GeV})^2$ are required to generate the required VEV of
$S$. Due to the small values of $\lambda$ and $\kappa$,
$A_\lambda$ and $A_\kappa$ remain small after the RG evolution from $M$
down to the weak scale, leading to a pseudo $R$-axion with a mass
$m_{a_1} \lsim 1$~ GeV. Now $a_1$ is particularly light since, for
small $\kappa$, it simultaneously plays the r\^ole of a Peccei-Quinn
pseudo Goldstone boson. However, due to the small value of $\lambda$,
the couplings of $a_1$ (with doublet components $\lsim 10^{-3}$) are
tiny, and this CP-odd scalar would be very hard to detect; the branching
ratios $h_i \to a_1 a_1$ are practically vanishing.

The CP-even Higgs sector is still compatible with LEP constraints if
$M$ is very large (and $M_{SUSY}$ somewhat larger than above), leading
to a sparticle spectrum (and $A_{t}$) in the 1~TeV range such that
top/stop induced radiative corrections lift up the CP-even Higgs
masses. Interestingly, in spite of $\lambda \sim 10^{-2}$, large values
for $\mu_{eff} = \lambda\left< S\right>$ still generate a large
singlet/doublet mixing for the two lightest CP-even scalars. As an
example, point P5 (which gives a satisfactory supersymmetric
contribution to the muon anomalous magnetic moment)  is shown in
Tables~1 and 2. $m_{h_1} \sim 94$~GeV is well below 114~GeV, but the
singlet component of $h_1$ is $\sim$ 88\% implying reduced couplings to
gauge bosons. The state $h_2$ with a mass $m_{h_2} \sim 120$~GeV has
still a singlet component $\sim$ 48\%. With the help of its nonsinglet
components, the detection of both states seems feasible at the ILC
\cite{ilc}.
Also, the lightest neutralino is a nearly pure singlino (with
nonsinglet components $\lsim 3\times 10^{-3}$), which would appear at
the end of sparticle decay cascades \cite{casc}.

\vspace{7mm}

Throughout this paper we have not addressed the issue of dark matter.
Clearly, within GMSB models the LSP is the gravitino, but heavy
remnants from the messenger sector can also contribute to the relic
density  \cite{dimgp,messdm}. Its evaluation would require assumptions
on the messenger/sequestered sector and the reheating temperature after
inflation, and is beyond the scope of the present work. On the other
hand, general considerations can possibly help to constrain the large
variety of different scenarios found here.

\section{Conclusions}

We have seen in this analysis that the NMSSM can solve the
$\mu$-problem in GMSB models in a phenomenologically acceptable way.
Our starting point was a derivation of the magnitude of all possible
supersymmetric and soft terms in a generalized NMSSM, that can be
radiatively generated by integrating out a sequestered/messenger sector
with couplings to the singlet superfield $\widehat{S}$. For the
phenomenological analysis, we confined ourselves to scenarios where
most of these terms are negligibly small. Nevertheless we found a large
variety of very different viable scenarios.

Scenarios with singlet tadpole terms $\it are$ acceptable, if the
linear terms in $\widehat{S}$ (or $S$) are generated to higher loop
order only, or if at least one small Yukawa coupling is involved. A
simple concrete model \cite{ue1995} with a direct coupling of
$\widehat{S}$ to the messengers is viable for a Yukawa coupling $\eta
\lsim 10^{-5}$. In the case of models with forbidden tadpole terms, as
those proposed in \cite{gr} and analysed in \cite{dgs}, we confirmed
the phenomenological viability observed in \cite{dgs} (at least for the
regions in parameter space without uncomfortably large values of
$A_\kappa$).

Quite interesting from the phenomenological point of view are the
scenarios with vanishing $A$-terms at the messenger scale: these
automatically lead to a light CP-odd Higgs scalar as studied in
\cite{raxion,lighthiggs}, which plays the r\^ole of a pseudo $R$-axion.
In view of the simplicity with which these scenarios can satisfy LEP
constraints, it would be very desirable to develop concrete models
which generate this structure for the effective NMSSM at the scale $M$.

Finally we recall that the Fortran routine NMGMSB, that allowed to 
obtain the results above, will be available on the website
\cite{nmssmtools}. With the help of corresponding input and output
files, further properties of the points P1 to P5 as sparticle masses,
couplings and branching ratios can be obtained.

\section*{Note added}

After the completion of this paper another viable scenario was proposed
in \cite{wagner}, in which the singlet does not couple to the
messenger/sequestered sector, but where the source of supersymmetry
breaking in the messenger sector is not SU(5) invariant.

\section*{Acknowledgements}

We are grateful to A. Djouadi and F. Domingo for helpful discussions,
and acknowledge support from the French ANR project PHYS@COL\&COS.

\section*{Appendix A}

In this appendix we summarize the expressions for the gaugino and
scalar masses (at the scale $M$), which are generated by gauge
mediation under the assumptions that the messenger sector involves
$n_5$ $(5 + \bar{5})$ representations under $SU(5)$ (additional $(10
+\overline{10})$ representations can be taken care of by adding three
units to $n_5$) with a common SUSY mass $M$, and $F$-type mass
splittings $m^2$ among the scalars and pseudoscalars. The $U(1)_Y$
coupling $\alpha_1$ is defined in the SM normalization (not in the GUT
normalization). For convenience we define the scale $M_{SUSY} = m^2/M$
and the parameter $x = M_{SUSY}/M$ (typically $\ll 1$). The required
one loop and two loop functions are \cite{dimgp,loopf}
\bea
f_1(x)&=&\frac{1}{x^2}\left((1+x)\ln(1+x)+(1-x)\ln(1-x)\right),\nn\\
f_2(x)&=&\frac{1+x}{x^2}\left(\ln(1+x)-2Li_2\left(\frac{x}{1+x}\right)
+\frac{1}{2}Li_2\left(\frac{2x}{1+x}\right)\right) + (x\to -x)\ ,\nn
\eea
which satisfy $f_1(x\to 0) = f_2(x\to 0) = 1$.

Then the gaugino masses are given by
\bea
M_1 &=& \frac{\alpha_1}{4\pi} M_{SUSY} f_1(x) \frac{5}{3}n_5\ ,\nn\\
M_2 &=& \frac{\alpha_2}{4\pi} M_{SUSY} f_1(x) n_5\ ,\nn\\
M_3 &=& \frac{\alpha_3}{4\pi} M_{SUSY} f_1(x) n_5\ ,\nn
\eea
and the scalar masses squared by
\bea
m^2 &=& \frac{M_{SUSY}^2}{16\pi^2}\left( \frac{10}{3}Y^2 \alpha_1^2
+\frac{3}{2} \alpha_2^2{^{(1)}} 
+\frac{8}{3} \alpha_3^2{^{(2)}} \right) f_2(x) n_5\ .\nn
\eea
The terms $^{(1)}$ are present for $SU(2)$ doublets only, and the terms
$^{(2)}$ for $SU(3)$ triplets only. The hypercharges $Y$ are
\begin{table}[!ht]
\begin{center}
\begin{tabular}{|l|r|r|r|r|r|r|}
\hline
 & $u_L/d_L$ & $u_R$ & $d_R$ & $\nu_L/e_L$ & $e_R$ & $H_u,H_d$\\
 \hline
$Y$ & $\frac{1}{6}$\phantom{x\Large$\frac{1}{1}$} & $\frac{2}{3}$ &
$-\frac{1}{3}$ &  $ \phantom{x}-\frac{1}{2}$ \phantom{x} & $-1$ &
$\pm\frac{1}{2} \phantom{xx}$\\
\hline
\end{tabular}
\end{center}
\end{table}

%\newpage

\section*{Appendix B}

In this appendix we give the one loop $\beta$-functions for the
parameters in the Higgs sector of the general NMSSM, defined by a
superpotential
\bea\nn
W &=& \lambda \widehat{S} \widehat{H_u} \widehat{H_d} + \frac{\kappa}{3} 
\widehat{S}^3 +\mu\widehat{H_u} \widehat{H_d} 
+\mu'\widehat{S}^2 +\xi_F\widehat{S} \\ \nn
&&+h_t \widehat{Q}_3\widehat{H}_u \widehat{T}^c_R 
-h_b \widehat{Q}_3\widehat{H}_d \widehat{B}^c_R 
-h_\tau \widehat{L}_3\widehat{H}_d \widehat{L}^c_R\nn
\eea
and soft terms
\bea\nn
V_{soft} &=& m_u^2 |H_u|^2 +m_d^2 |H_d|^2 +m_S^2|S|^2
+(\lambda A_\lambda S H_u H_d +\frac{\kappa}{3}
A_\kappa S^3 + m_3^2 H_u H_d +m_s'^2 S^2 +\xi_S S \\ \nn
&& + h_tA_t Q_3H_u T^c_R -h_bA_b Q_3H_d B^c_R 
-h_\tau A_\tau L_3H_d L^c_R
 + h.c.)\ ,
\eea
under the assumption $\sum_i Y_i m_i^2 = 0$, which is always satisfied
for GMSB models.

\bea\nn
\frac{d\lambda^2}{d\ln Q^2} &=&
\frac{\lambda^2}{16\pi^2} \left(4\lambda^2 +2 \kappa^2 + 3 (h_t^2
+h_b^2) +h_\tau^2 -g_1^2 -3g_2^2 \right)\\ \nn
\frac{d\kappa^2}{d\ln Q^2} &=&
\frac{\kappa^2}{16\pi^2} \left(6\lambda^2 +6\kappa^2\right)\\ \nn
\frac{d h_t^2}{d\ln Q^2} &=&
\frac{ h_t^2}{16\pi^2} \left(\lambda^2 +6h_t^2 +h_b^2-\frac{16}{3}g_3^2
-3g_2^2 -\frac{13}{9}g_1^2\right)\\ \nn
\frac{d h_b^2}{d\ln Q^2} &=&
\frac{h_b^2}{16\pi^2} \left(\lambda^2 +6h_b^2 +h_t^2 +h_\tau^2
-\frac{16}{3}g_3^2 -3g_2^2 -\frac{7}{9}g_1^2\right)\\ \nn
\frac{d h_\tau^2}{d\ln Q^2} &=&
\frac{h_\tau^2}{16\pi^2} \left(\lambda^2 +3h_b^2 +4h_\tau^2
-3g_2^2 -3g_1^2\right)\\ \nn
\frac{d\mu}{d\ln Q^2} &=&
\frac{\mu}{16\pi^2} \left(\lambda^2 +\frac{3}{2}(h_t^2 +h_b^2)
+\frac{1}{2}h_\tau^2 -\frac{1}{2}(g_1^2+3g_2^2)\right)\\ \nn
\frac{d\mu'}{d\ln Q^2} &=&
\frac{\mu'}{16\pi^2} \left(2\lambda^2 +2\kappa^2\right)\\ \nn
\frac{d\xi_F}{d\ln Q^2} &=&
\frac{\xi_F}{16\pi^2} \left(\lambda^2 +\kappa^2\right)\\ \nn
\frac{d A_\lambda}{d\ln Q^2} &=&
\frac{1}{16\pi^2} \left(4\lambda^2 A_\lambda +2\kappa^2A_\kappa
+3(h_t^2A_t+h_b^2A_b) +h_\tau^2A_\tau+g_1^2M_1+3g_2^2M_2\right)\\ \nn
\frac{d A_\kappa}{d\ln Q^2} &=&
\frac{1}{16\pi^2} \left(6\lambda^2 A_\lambda
+6\kappa^2A_\kappa\right)\\ \nn
\frac{d A_t}{d\ln Q^2} &=&
\frac{1}{16\pi^2} \left(\lambda^2 A_\lambda +6h_t^2 A_t +h_b^2A_b
+\frac{13}{9}g_1^2M_1+3g_2^2M_2 +\frac{16}{3}g_3^2M_3\right)\\ \nn
\frac{d A_b}{d\ln Q^2} &=&
\frac{1}{16\pi^2} \left(\lambda^2 A_\lambda +6h_b^2 A_b +h_t^2A_t
+h_\tau^2A_\tau
+\frac{7}{9}g_1^2M_1+3g_2^2M_2 +\frac{16}{3}g_3^2M_3\right)\\ \nn
\frac{d A_\tau}{d\ln Q^2} &=&
\frac{1}{16\pi^2} \left(\lambda^2 A_\lambda +3h_b^2 A_b
+4h_\tau^2A_\tau +3g_1^2M_1+3g_2^2M_2 \right)\\ \nn
\frac{d m_u^2}{d\ln Q^2} &=&
\frac{1}{16\pi^2} \left(\lambda^2 (m_u^2+m_d^2+m_S^2+A_\lambda^2)
 +3h_t^2 (m_u^2+m_T^2+m_Q^2+A_t^2) +\frac{g_1^2}{2}(m_u^2-m_d^2)
 \right. \\ \nn
&&- \left. g_1^2M_1^2-3g_2^2M_2^2 \right)\\ \nn
\frac{d m_d^2}{d\ln Q^2} &=&
\frac{1}{16\pi^2}\left(\lambda^2 (m_u^2+m_d^2+m_S^2+A_\lambda^2)
 +3h_b^2 (m_d^2+m_B^2+m_Q^2+A_b^2) \right. \\ \nn
&& \left. +h_\tau^2(m_d^2+m_\tau^2+m_L^2+A_\tau^2) 
 -\frac{g_1^2}{2}(m_u^2-m_d^2)
-g_1^2M_1^2-3g_2^2M_2^2 \right) \\ \nn
\frac{d m_S^2}{d\ln Q^2} &=&
\frac{1}{16\pi^2}\left(2\lambda^2 (m_u^2+m_d^2+m_S^2+A_\lambda^2)
 +\kappa^2 (6m_S^2+2A_\kappa^2) \right) \\ \nn
 \frac{d m_3^2}{d\ln Q^2} &=&
\frac{1}{16\pi^2}\left(\frac{m_3^2}{2}(6\lambda^2+3h_t^2+3h_b^2
+h_\tau^2-g_1^2-3g_2^2) +2\lambda\kappa m_S'^2 \right. \\ \nn
&&\left. + \mu(2\lambda^2A_\lambda -3h_t^2A_t -3h_b^2A_b 
-h_\tau^2A_\tau +g_1^2M_1 +3g_2^2M_2) \right) \\ \nn
 \frac{d m_S'^2}{d\ln Q^2} &=&
\frac{1}{16\pi^2}\left(m_S'^2(2\lambda^2+4\kappa^2) 
+2\lambda\kappa m_3^2 
+4\mu'(\lambda^2A_\lambda +\kappa^2A_\kappa) \right) \\ \nn
\frac{d \xi_S}{d\ln Q^2} &=&
\frac{1}{16\pi^2}\left(\xi_S(\lambda^2+\kappa^2) 
+2\lambda\mu(m_u^2+m_d^2) +4\kappa \mu' m_S^2
+2\lambda m_3^2(A_\lambda+2\mu') \right. \\ \nn
&&\left. +2\xi_F(\lambda^2A_\lambda +\kappa^2A_\kappa
+2\kappa m_S'^2(A_\kappa+2\mu') \right) 
\eea

%\newpage


\begin{thebibliography}{99}

\bibitem{early1} M.~Dine, W.~Fischler and M.~Srednicki,
  %``Supersymmetric Technicolor,''
  Nucl.\ Phys.\  B {\bf 189} (1981) 575;\\
S.~Dimopoulos and S.~Raby,
  %``Supercolor,''
  Nucl.\ Phys.\  B {\bf 192} (1981) 353;\\
M.~Dine and M.~Srednicki,
  %``More Supersymmetric Technicolor,''
  Nucl.\ Phys.\  B {\bf 202} (1982) 238.
  
\bibitem{earlyoraif}
M.~Dine and W.~Fischler,
  %``A Phenomenological Model Of Particle Physics Based On 
  %Supersymmetry,''
  Phys.\ Lett.\  B {\bf 110} (1982) 227;\\
M.~Dine and W.~Fischler,
  %``A Supersymmetric Gut,''
  Nucl.\ Phys.\  B {\bf 204} (1982) 346;\\
L.~Alvarez-Gaume, M.~Claudson and M.~B.~Wise,
  %``Low-Energy Supersymmetry,''
  Nucl.\ Phys.\  B {\bf 207} (1982) 96;\\
S.~Dimopoulos and S.~Raby,
  %``Geometric Hierarchy,''
  Nucl.\ Phys.\  B {\bf 219} (1983) 479.
  
\bibitem{des}
N.~Dragon, U.~Ellwanger and M.~G.~Schmidt,
  %``Sliding Scales In Minimal Supergravity,''
  Phys.\ Lett.\  B {\bf 145} (1984) 192;
%U.~Ellwanger, N.~Dragon and M.~G.~Schmidt,
  %``Hierarchies From Minimal Supergravity,''
  Nucl.\ Phys.\  B {\bf 255} (1985) 549;
%N.~Dragon, U.~Ellwanger and M.~G.~Schmidt,
  %``Radiative Scale Fixing And The Mass Spectrum In Models Based 
  %On SU(N,1) Minimal Supergravity,''
  Phys.\ Lett.\  B {\bf 154} (1985) 373;
%U.~Ellwanger, N.~Dragon and M.~G.~Schmidt,
  %``A Grand Unified Theory Based On SU(N,1) Minimal Supergravity,''
  Z.\ Phys.\  C {\bf 29} (1985) 209.
  
\bibitem{ue1995}
 U.~Ellwanger,
  %``How To Limit Radiative Corrections To The Cosmological Constant 
  %By M(SUSY)**4,''
  Phys.\ Lett.\  B {\bf 349} (1995) 57
  [arXiv:hep-ph/9501227].
 
\bibitem{dsb}
I.~Affleck, M.~Dine and N.~Seiberg,
  %``Dynamical Supersymmetry Breaking In Supersymmetric QCD,''
  Nucl.\ Phys.\  B {\bf 241} (1984) 493;
%I.~Affleck, M.~Dine and N.~Seiberg,
  %``Dynamical Supersymmetry Breaking In Chiral Theories,''
  Phys.\ Lett.\  B {\bf 137} (1984) 187;
%I.~Affleck, M.~Dine and N.~Seiberg,
  %``Exponential Hierarchy From Dynamical Supersymmetry Breaking,''
  Phys.\ Lett.\  B {\bf 140} (1984) 59;
%I.~Affleck, M.~Dine and N.~Seiberg,
  %``Dynamical Supersymmetry Breaking In Four-Dimensions And Its
  %Phenomenological Implications,''
  Nucl.\ Phys.\  B {\bf 256} (1985) 557;\\
Y.~Meurice and G.~Veneziano,
  %``SUSY Vacua Versus Chiral Fermions,''
  Phys.\ Lett.\  B {\bf 141} (1984) 69.
  
\bibitem{dsbnmssm}
M.~Dine and A.~E.~Nelson,
  %``Dynamical supersymmetry breaking at low-energies,''
  Phys.\ Rev.\  D {\bf 48} (1993) 1277
  [arXiv:hep-ph/9303230];\\
M.~Dine, A.~E.~Nelson and Y.~Shirman,
  %``Low-Energy Dynamical Supersymmetry Breaking Simplified,''
  Phys.\ Rev.\  D {\bf 51} (1995) 1362
  [arXiv:hep-ph/9408384];\\
 M.~Dine, A.~E.~Nelson, Y.~Nir and Y.~Shirman,
  %``New tools for low-energy dynamical supersymmetry breaking,''
  Phys.\ Rev.\  D {\bf 53} (1996) 2658
  [arXiv:hep-ph/9507378].
  
\bibitem{iss}
  K.~Intriligator, N.~Seiberg and D.~Shih,
  %``Dynamical SUSY breaking in meta-stable vacua,''
  JHEP {\bf 0604} (2006) 021
  [arXiv:hep-th/0602239].
  
\bibitem{nmssm1}
P. Fayet, Nucl. Phys. B \textbf{90} (1975) 104; Phys. Lett. B
\textbf{64} (1976) 159; Phys. Lett. B \textbf{69} (1977) 489 and Phys.
Lett. B \textbf{84} (1979) 416.

\bibitem{nmssm2}
H.P. Nilles, M. Srednicki and D. Wyler, Phys. Lett. B \textbf{120}
(1983) 346;\\ 
J.M. Frere, D.R. Jones and S. Raby, Nucl. Phys. B \textbf{222} (1983)
11.

\bibitem{gm}
  G.~F.~Giudice and A.~Masiero,
  %``A Natural Solution to the mu Problem in Supergravity Theories,''
  Phys.\ Lett.\  B {\bf 206} (1988) 480.
  
\bibitem{hmz}
A.~Hebecker, J.~March-Russell and R.~Ziegler,
  ``Inducing the $\mu$ and the $B\mu$ Term by the Radion and the 5d
  Chern-Simons Term,''
  arXiv:0801.4101 [hep-ph].

\bibitem{tadpole}
J.~Polchinski and L.~Susskind,
  %``Breaking Of Supersymmetry At Intermediate-Energy,''
  Phys.\ Rev.\  D {\bf 26} (1982) 3661;\\
H.~P.~Nilles, M.~Srednicki and D.~Wyler,
  %``Constraints On The Stability Of Mass Hierarchies In Supergravity,''
  Phys.\ Lett.\  B {\bf 124} (1983) 337;\\
A.~B.~Lahanas,
  %``Light Singlet, Gauge Hierarchy And Supergravity,''
  Phys.\ Lett.\  B {\bf 124} (1983) 341;\\
U.~Ellwanger,
  %``Nonrenormalizable Interactions From Supergravity, Quantum Corrections And
  %Effective Low-Energy Theories,''
  Phys.\ Lett.\  B {\bf 133} (1983) 187;\\
J.~Bagger and E.~Poppitz,
  %``Destabilizing divergences in supergravity coupled supersymmetric
  %theories,''
  Phys.\ Rev.\ Lett.\  {\bf 71} (1993) 2380  [arXiv:hep-ph/9307317];\\
 J.~Bagger, E.~Poppitz and L.~Randall,
  %``Destabilizing divergences in supergravity theories at two loops,''
  Nucl.\ Phys.\  B {\bf 455} (1995) 59
  [arXiv:hep-ph/9505244].
  
\bibitem{neme} D.~Nemeschansky,
  %``The Sliding Singlet,''
  Nucl.\ Phys.\  B {\bf 234} (1984) 379.
  
\bibitem{dgp}
G.~R.~Dvali, G.~F.~Giudice and A.~Pomarol,
  %``The $\mu$-Problem in Theories with Gauge-Mediated Supersymmetry Breaking,''
  Nucl.\ Phys.\  B {\bf 478} (1996) 31
  [arXiv:hep-ph/9603238].

\bibitem{gmnm1}
P.~Ciafaloni and A.~Pomarol,
  %``Dynamical determination of the supersymmetric Higgs mass,''
  Phys.\ Lett.\  B {\bf 404} (1997) 83
  [arXiv:hep-ph/9702410];\\
K.~Agashe and M.~Graesser,
  %``Improving the fine tuning in models of low energy gauge mediated
  %supersymmetry breaking,''
  Nucl.\ Phys.\  B {\bf 507} (1997) 3
  [arXiv:hep-ph/9704206].
  
\bibitem{gr}  
G.~F.~Giudice and R.~Rattazzi,
  %``Extracting supersymmetry-breaking effects from wave-function
  %renormalization,''
  Nucl.\ Phys.\  B {\bf 511} (1998) 25
  [arXiv:hep-ph/9706540].
  
\bibitem{gmnm2}
A.~de Gouvea, A.~Friedland and H.~Murayama,
  Phys.\ Rev.\  D {\bf 57} (1998) 5676
  [arXiv:hep-ph/9711264];\\
T.~Han, D.~Marfatia and R.~J.~Zhang,
  Phys.\ Rev.\  D {\bf 61} (2000) 013007
  [arXiv:hep-ph/9906508];\\
Z.~Chacko and E.~Ponton,
  %``Yukawa deflected gauge mediation,''
  Phys.\ Rev.\  D {\bf 66} (2002) 095004
  [arXiv:hep-ph/0112190];\\
M.~Dine and J.~Mason,
  ``Dynamical Supersymmetry Breaking and Low Energy Gauge Mediation,''
  arXiv:0712.1355 [hep-ph].
  
\bibitem{dgs}  
A.~Delgado, G.~F.~Giudice and P.~Slavich,
  %``Dynamical mu Term in Gauge Mediation,''
  Phys.\ Lett.\  B {\bf 653} (2007) 424
  [arXiv:0706.3873 [hep-ph]].
  
\bibitem{gkr}
G.~F.~Giudice, H.~D.~Kim and R.~Rattazzi,
  ``Natural mu and Bmu in gauge mediation,''
  arXiv:0711.4448 [hep-ph].
 
\bibitem{slha2}
  B.~C.~Allanach {\it et al.},
  ``SUSY Les Houches Accord 2,''
  arXiv:0801.0045 [hep-ph].
  
\bibitem{nmssmtools}
  U.~Ellwanger, J.~F.~Gunion and C.~Hugonie,
  JHEP {\bf 0502} (2005) 066
  [arXiv:hep-ph/0406215];\\
  U.~Ellwanger and C.~Hugonie,
  Comput.\ Phys.\ Commun.\  {\bf 177} (2007) 399,\\
  {\sf http://www.th.u-psud.fr/NMHDECAY/nmssmtools.html}

\bibitem{lhg}
 S.~Schael {\it et al.}  [ALEPH, DELPHI, L3 and OPAL Collaborations],
  Eur.\ Phys.\ J.\  C {\bf 47} (2006) 547

\bibitem{bphys}
F.~Domingo and U.~Ellwanger,
  %``Updated Constraints from B Physics on the MSSM and the NMSSM,''
  JHEP {\bf 0712} (2007) 090
  [arXiv:0710.3714 [hep-ph]].
 
\bibitem{nnmssm} 
  C.~Panagiotakopoulos and A.~Pilaftsis,
  Phys.\ Rev.\  D {\bf 63} (2001) 055003   [arXiv:hep-ph/0008268];\\
    A.~Dedes, C.~Hugonie, S.~Moretti and K.~Tamvakis,
  Phys.\ Rev.\  D {\bf 63} (2001) 055009 [arXiv:hep-ph/0009125];\\
    A.~Menon, D.~E.~Morrissey and C.~E.~M.~Wagner,
  %``Electroweak baryogenesis and dark matter in the nMSSM,''
  Phys.\ Rev.\  D {\bf 70} (2004) 035005
  [arXiv:hep-ph/0404184].
  
\bibitem{adjk}
  S.~A.~Abel, C.~Durnford, J.~Jaeckel and V.~V.~Khoze,
  ``Patterns of Gauge Mediation in Metastable SUSY Breaking,''
  arXiv:0712.1812 [hep-ph].
  
\bibitem{adjk1}
  S.~Abel, C.~Durnford, J.~Jaeckel and V.~V.~Khoze,
  ``Dynamical breaking of $U(1)_{R}$ and supersymmetry in a metastable
  vacuum,'' arXiv:0707.2958 [hep-ph].
  
\bibitem{raxion}
B.~A.~Dobrescu, G.~Landsberg and K.~T.~Matchev, Phys.\ Rev.\ D {\bf 63}
  (2001) 075003 [arXiv:hep-ph/0005308];\\
B.~A.~Dobrescu and K.~T.~Matchev, JHEP {\bf 0009} (2000) 031 
  [arXiv:hep-ph/0008192];\\
R.~Dermisek and J.~F.~Gunion,
  Phys.\ Rev.\  D {\bf 75} (2007) 075019
  [arXiv:hep-ph/0611142].

\bibitem{lighthiggs}
R.~Dermisek and J.~F.~Gunion,
  Phys.\ Rev.\ Lett.\  {\bf 95} (2005) 041801 [arXiv:hep-ph/0502105] and
  Phys.\ Rev.\ D {\bf 73} (2006) 111701 
  [arXiv:hep-ph/0510322];\\
U.~Ellwanger, J.~F.~Gunion and C.~Hugonie, JHEP {\bf 0507} (2005) 041 
 [arXiv:hep-ph/0503203];\\
S.~Chang, P.~J.~Fox and N.~Weiner,
  JHEP {\bf 0608} (2006) 068 [arXiv:hep-ph/0511250];\\
P.~W.~Graham, A.~Pierce and J.~G.~Wacker, ``Four taus at the
  Tevatron,'' arXiv:hep-ph/0605162;\\
S.~Moretti, S.~Munir and P.~Poulose, Phys.\ Lett.\ B {\bf 644} (2007)
  241  [arXiv:hep-ph/0608233];\\
S.~Chang, P.~J.~Fox and N.~Weiner,
  Phys.\ Rev.\ Lett.\  {\bf 98} (2007) 111802 [arXiv:hep-ph/0608310];\\
T.~Stelzer, S.~Wiesenfeldt and S.~Willenbrock,
  Phys.\ Rev.\  D {\bf 75} (2007) 077701 [arXiv:hep-ph/0611242];\\
U.~Aglietti {\it et al.}, ``Tevatron-for-LHC report: Higgs,''
  arXiv:hep-ph/0612172;\\
E.~Fullana and M.~A.~Sanchis-Lozano,
  Phys.\ Lett.\  B {\bf 653} (2007) 67
  [arXiv:hep-ph/0702190];\\  
K.~Cheung, J.~Song and Q.~S.~Yan,
  Phys.\ Rev.\ Lett.\  {\bf 99} (2007) 031801 [arXiv:hep-ph/0703149];\\
M.~A.~Sanchis-Lozano,
  ``A light non-standard Higgs boson: to be or not to be at a (Super) B
  factory?,'' arXiv:0709.3647 [hep-ph];\\
M.~Carena, T.~Han, G.~Y.~Huang and C.~E.~M.~Wagner, 
   arXiv:0712.2466 [hep-ph];\\ 
J.R. Forshaw, J.F. Gunion, L. Hodgkinson, A. Papaefstathiou and  A.D.
   Pilkington, arXiv:0712.3510 [hep-ph];\\
Z.~Heng, R.~J.~Oakes, W.~Wang, Z.~Xiong and J.~M.~Yang,
  ``B meson Dileptonic Decays in NMSSM with a Light CP-odd Higgs Boson,''
  arXiv:0801.1169 [hep-ph];\\
A.~Djouadi {\it et al.},
  ``Benchmark scenarios for the NMSSM,''
  arXiv:0801.4321 [hep-ph].
 
\bibitem{ilc} U.~Ellwanger, J.~F.~Gunion, C.~Hugonie and S.~Moretti,
  ``Towards a no-lose theorem for NMSSM Higgs discovery at the LHC,''
  arXiv:hep-ph/0305109, published in  G.~Weiglein {\it et al.}  
  [LHC/LC Study Group],
  Phys.\ Rept.\  {\bf 426} (2006) 47
  [arXiv:hep-ph/0410364].

\bibitem{casc}
U.~Ellwanger and C.~Hugonie,
  Eur.\ Phys.\ J.\  C {\bf 5} (1998) 723 [arXiv:hep-ph/9712300] and
  Eur.\ Phys.\ J.\  C {\bf 13} (2000) 681 [arXiv:hep-ph/9812427];\\
   V.~Barger, P.~Langacker and G.~Shaughnessy,
  %``Neutralino signatures of the singlet extended MSSM,''
  Phys.\ Lett.\  B {\bf 644} (2007) 361
  [arXiv:hep-ph/0609068].

\bibitem{dimgp}
 S.~Dimopoulos, G.~F.~Giudice and A.~Pomarol, Phys.\ Lett.\  B {\bf
  389}, 37 (1996) [arXiv:hep-ph/9607225].
 
\bibitem{messdm}  
T.~Han and R.~Hempfling,
  %``Messenger sneutrinos as cold dark matter,''
  Phys.\ Lett.\  B {\bf 415} (1997) 161
  [arXiv:hep-ph/9708264];\\
K.~Jedamzik, M.~Lemoine and G.~Moultaka, Phys.\ Lett.\  B {\bf 645}
  (2007) 222 [arXiv:hep-ph/0504021] and   Phys.\ Rev.\  D {\bf 73}
  (2006) 043514 [arXiv:hep-ph/0506129].

\bibitem{loopf}
  S.~P.~Martin, Phys.\ Rev.\  D {\bf 55} (1997) 3177
  [arXiv:hep-ph/9608224].

\bibitem{wagner}
  T.~Liu and C.~E.~M.~Wagner,
  ``Dynamically Solving the $\mu/B_\mu$ Problem in Gauge-mediated
  Supersymmetry Breaking,'' arXiv:0803.2895 [hep-ph].

\end{thebibliography}
\end{document}